\documentclass[12pt]{article}
\usepackage{amssymb,amsmath,epsfig}
\usepackage{graphicx}
\allowdisplaybreaks

\begin{document}

\title{\bf Decoupled Quark Stars Relativistic Models in the Regime of Self-interacting Brans-Dicke Gravity}

\author{M. Sharif$^1$ \thanks{msharif.math@pu.edu.pk} and Amal Majid$^2$
\thanks{amalmajid89@gmail.com, amal.majid@ucp.edu.pk}\\
$^1$ Department of Mathematics and Statistics, The University of Lahore,\\
1-KM Defence Road Lahore, Pakistan.\\
$^2$ Department of Mathematics, University of Central Punjab,\\
Lahore, Pakistan.}

\date{}

\maketitle
\begin{abstract}
In this work, we generate two static anisotropic solutions for a
sphere containing quark matter in the framework of self-interacting
Brans-Dicke theory. For this purpose, we add an anisotropic source
in the seed distribution and decouple the field equations through
deformation in the radial metric function. As a result of this
transformation, the field equations are disintegrated into two
systems which separately include the effects of isotropic and
anisotropic sources. The system related to the additional source is
solved via the MIT bag model equation of state. We consider Tolman V
spacetime and Karmarkar condition to formulate two solutions for the
isotropic sector which are extended to the anisotropic domain via
decoupling technique. The junction conditions at the boundary
determine the unknown parameters in terms of mass and radius of the
spherical object. We investigate the viability and stability of the
constructed strange star models in the presence of massive scalar
field corresponding to three strange star candidates: Her X-1, PSR
J1614-2230 and 4U 1608-52. It is concluded that the anisotropic
models are well-behaved as they fulfill the necessary requirements
for lower as well as higher values of the decoupling parameter.
\end{abstract}

\section{Introduction}

The general theory of relativity is modified to incorporate the
effects of cosmic expansion. In this regard, the family of
scalar-tensor theories has gained attention. Brans-Dicke (BD) theory
(a scalar-tensor theory prototype) includes a spacetime varying
scalar field $\vartheta=\frac{1}{G}.$ ($G$ is Newton's gravitational
constant) along with the metric tensor to describe the effects of
gravitational field \cite{18}. A BD coupling parameter
($\omega_{BD}$) couples the matter distribution and the massless
scalar field. Different values of the tunable coupling parameter
explain different cosmological scenarios. However, a conflict occurs
when large values of $\omega_{BD}$ in the weak-field limit \cite{20}
do not coincide with the small values required to describe the
inflationary era \cite{19}. In order to remove the discrepancy, a
potential function $V(\varsigma)$ is introduced which allots a mass
to the scalar field ($\varsigma$). This modification leads to
self-interacting BD (SBD) theory which admits all values of the
coupling parameter greater than $-\frac{3}{2}$ for
$m_\varsigma>2\times10^{-25}GeV$ ($m_\varsigma$ denotes mass of the
scalar field) \cite{23'}.

Further, a self-interacting scalar field with a non-vanishing
potential supports the current cosmological expansion. On the other
hand, the case of the quadratic field Lagrangian corresponds to the
case of a dynamical cosmological constant. Thus, in cosmological
scenarios, many efforts have been made to inspect the dynamical
behavior of the models dependent on the choice of the non-vanishing
potential \cite{1119, 1119a}. The situation becomes more intriguing
when issues from string theories are taken into account which
motivates the testing of simple asymptotically anti-de Sitter models
corresponding to potentials having non-standard behavior. Thus,
self-interacting field with a non-trivial potential are key
ingredients in cosmological models and singularities.

In the context of the unsolved cosmic censorship issue, the scalar
field sources have usually been considered as free, i.e., they are
minimally coupled to gravity via equivalence principle. The
existence of naked singularities has been shown for free scalar
fields. However, such singularities are non-generic corresponding to
the choice of the initial data \cite{1111}. In recent years,
gravitational collapse of homogeneous self-interacting scalar fields
has been studied and the formation of naked singularities has been
found by choosing special potentials \cite{1112, 1112a}. All in all,
in both scenarios (cosmological as well as astrophysical) it is
interesting to employ self-interacting scalar field dynamics to
construct stellar models such as quark stars and to predict their
qualitative behavior.

Different solutions in astrophysical scenario have been devised in
BD as well as SBD theory. Buchdahl \cite{22'} employed spherical as
well as axial vacuum solutions of general relativity (GR) to compute
a corresponding family of solutions in BD theory. Bruckman and Kazes
\cite{24} adopted a linear relation between pressure and density to
formulate an exact spherical solution with perfect fluid. A complex
coordinate transformation was applied to obtain a Demianski-type
metric by Krori and Bhattacharjee \cite{25a}. The effect of scalar
field has also been extensively discussed on slowly as well as
rapidly rotating neutron stars \cite{55a}-\cite{55d}.

Self-gravitating objects (such as stellar structures, galaxies and
planets) contain significant clues regarding the history and
structure of the vast universe. Stars are one of the most widely
studied self-gravitating objects. These astronomical bodies,
composed of accumulated dust and hydrogen gas, support themselves
against inward gravity through fusion reactions. When stars exhaust
the limited supply of fuel in their cores, they collapse to one of
the three compact structures: white dwarfs, neutron stars or black
holes. Neutron stars are intriguing cosmic objects with a mass of
1$M_{\bigodot}$ to 3$M_{\bigodot}$ ($M_{\bigodot}$ denotes solar
mass). The degeneracy pressure in their interior stops further
gravitational collapse. However, it is hypothesized that if the mass
of a neutron star exceeds the limit of 3$M_{\bigodot}$, it continues
its inward collapse leading to extreme temperature and pressure
within the core. Consequently, neutrons breakdown into constituent
elements known as quarks which lead to the formation of strange
quark stars. It has been proposed that quark matter formed during
the transition of a neutron star to a strange star maintains its
stability \cite{8a}.

Bodmer \cite{99} and Witten \cite{100} were among the first who
speculated the existence of compact stars consisting of deconfined
quarks made up of the three lightest quark flavor states. It was
recently conjectured that the star RX J 1856.5A-3754 could fit the
description of a quark star \cite{108}. However, this statement was
heavily criticized because of the speculated radius used in the
study. Thus, the existence of quark stars is still an open question.
Later Annala et al. \cite{109} considered the possibility of stellar
cores filled with quark matter and found that the characteristics of
matter distribution in the cores of neutron stars favored the
existence of strange stars. Moreover, researchers believe that the
enormous release of radiation and energy in extremely luminous
supernovae can be explained through quark stars \cite{101, 101a}.

Strange quark stars are smaller than neutron stars and possess
extremely dense cores. In order to describe these ultra-dense
structures, a suitable equation of state (EoS) encompassing the
essential features of quark-gluon plasma, is required. A model that
adequately describes the interactions of quarks at high density
levels is still under investigation. However, in the absence of the
best fit, researchers have used MIT bag model EoS to examine
different properties of strange quark stars \cite{102}. This model
distinguishes true vacuum (global minimum of energy in a stable
setup) from false vacuum (local minimum of energy in an unstable
setup) through a bag constant ($\mathcal{B}$). Moreover, the
statistical data obtained from GW170817 \cite{104} and GW190425
\cite{105} was used to predict the compactness and mass of quark
stars. These predictions match the estimates made via the MIT bag
model which strengthens the choice of this EoS. Bhar \cite{111} used
the MIT bag model to compute a solution representing strange stars
in the background of Krori and Barua spacetime. The bag model has
also been used to investigate the stability of quark stars by
introducing radial perturbations \cite{112}. Deb et al. \cite{113}
adopted the EoS to construct a charged stellar structure. A complex
spherical model consisting of three layers was developed by employing
MIT bag model to represent the quark core \cite{204}. Recently,
MIT bag model was employed to examine the behavior of strange quark
stars under the effect of a self-interacting scalar field \cite{129,
129a}.

Field equations play a crucial role in determining the relation
between geometry and the matter source of a structure. Thus, in
order to explore the salient characteristics of astrophysical
systems, solutions of the field equations are required.
Schwarzschild \cite{1} formulated the first spherical solution of
the complex differential equations for a simple isotropic
configuration. However, studies suggest that the profile of cosmic
systems is intricate and complex. For instance, the compactness of
stellar remnants restricts the movement of particles in tangential
and radial directions. Thus, anisotropy in pressure is a vital
feature of these dense celestial bodies. Other processes like pion
condensation \cite{6} or the presence of super fluid \cite{4} may
give rise to anisotropy as well. Herrera and Santos \cite{7}
determined the source and impact of anisotropy on different features
(such as mass, mass-radius relation, stability, etc.) of stellar
objects. Researchers have formulated anisotropic models via
different techniques \cite{8}-\cite{206}. However, the task of
obtaining well-behaved spacetimes corresponding to realistic models
is still under discussion.

Recently, Ovalle simplified the extraction of solutions by proposing
the gravitational decoupling approach via minimal geometric
deformation (MGD) \cite{10}. In this scheme, an extra source is
successively added to a seed distribution (vacuum or isotropic) to
incorporate intricate features of physical systems. The degree of
freedom (due to larger number of unknowns) in the field equations is
reduced by splitting the field equations into two arrays via a
linear transformation in the radial metric function. These sets
exclusively include the effects of seed and additional gravitational
sources. Moreover, no energy is transferred between the two matter
distributions as they are individually conserved. The respective
solutions of these systems are combined to formulate a new solution.
An extension of this technique has also been proposed which induces
transformations in temporal as well as radial metric functions
\cite{28a}.

The efficient technique of gravitational decoupling has successfully
been employed to generate anisotropic solutions from well-known
ansatz. The anisotropic extensions of Tolman IV have been obtained
in braneworld \cite{11} as well as GR \cite{13} via MGD scheme.
Isotropic configurations corresponding to Durgapal-Fuloria
\cite{15}, Krori-Barua \cite{33'}, Heintzmann \cite{16} and
Schwarzschild \cite{33b} interior solutions have been decoupled to
evaluate their corresponding anisotropic versions. Contreras
\cite{33a, 33c} deformed the metric component of (2+1)- dimensional
spacetime to decouple the field equations with a cosmological
constant. Singh et al. \cite{12} derived an anisotropic spacetime by
incorporating the embedding class-one condition in the decoupling
approach. Hensh and Stuchlik \cite{16a} constructed an anisotropic
counterpart of Tolman VII by the method of gravitational decoupling.
Sharif and Ama-Tul-Mughani \cite{17a,17b} utilized the MGD scheme to
extend charged solutions to anisotropic domain in a cloud of
strings. Anisotropic cosmological solutions have also been obtained
by applying the MGD technique to FLRW as well as Kantowski-Sachs
spacetimes \cite{36}. Pant et al. \cite{202} formulated charged anisotropic
version of Heintzmann solution via the same technique. Recently,
Contreras et al. \cite{36a} studied
axially symmetric black holes and showed that gravitational
decoupling can be adopted beyond spherical symmetry. Some
well-behaved solutions in SBD gravity as well as modified theories
have been computed through gravitational decoupling technique
\cite{28}-\cite{28f}.
In this paper, we formulate two anisotropic strange star models by
implementing MGD scheme on isotropic distributions in the context of
SBD gravity. For this purpose, we adopt Tolman V ansatz as well as
generate a spacetime via embedding condition. We use the MIT bag
model to investigate the viability and stability of the constructed
structures. We evaluate the field equations under the effect of the
additional source in section \textbf{2}. We deform the radial metric
function to split the field equations in section \textbf{3}. The
anisotropic extensions are checked for viability and stability in
section \textbf{4}. In section \textbf{5}, the important results are
discussed.

\section{Self-interacting Brans-Dicke Theory}

The SBD gravity is described by the action
\begin{equation}\label{0}
S=\int\sqrt{-g}(\mathcal{R}\varsigma-\frac{\omega_{BD}}{\varsigma}\nabla^{\gamma}\varsigma
\nabla_{\gamma}\varsigma
-V(\varsigma)+\emph{L}_m+\sigma\emph{L}_\Theta)d^{4}x,
\end{equation}
where relativistic units have been used. Here, $\mathcal{R}$ and
$\emph{L}_m$ denote the Ricci scalar and matter lagrangian whereas
$g=|g_{\gamma\delta}|$. The contribution of the anisotropic source
($\Theta$) via the lagrangian $\emph{L}_\Theta$ is controlled by the
decoupling parameter $\sigma$. The energy-momentum tensors of the
original and additional sources are respectively, written as
\begin{eqnarray*}
T_{\gamma\delta^{(m)}}&=&-\frac{2}{\sqrt{-g}}\frac{\delta(\sqrt{-g}\emph{L}_{m})}{\delta
g^{\gamma\delta}},\\
\Theta_{\gamma\delta}&=&-\frac{2}{\sqrt{-g}}\frac{\delta(\sqrt{-g}\emph{L}_{\Theta})}{\delta
g^{\gamma\delta}}.
\end{eqnarray*}
The additional gravitational source can be categorized as a scalar,
vector or tensor field. The field and wave equations derived from
the action (\ref{0}) are written as
\begin{eqnarray}\label{1}
G_{\gamma\delta}&=&T^{\text{(eff)}}_{\gamma\delta}=\frac{1}{\varsigma}(T_{\gamma\delta}^{(m)}+\sigma\Theta_{\gamma\delta}
+T_{\gamma\delta}^\varsigma),\\\label{2}
\Box\varsigma&=&\frac{T^{(m)}+\Theta}{3+2\omega_{BD}}+\frac{1}{3+2\omega_{BD}}
(\varsigma\frac{dV(\varsigma)}{d\varsigma}-2V(\varsigma)),
\end{eqnarray}
where
$\Box\varsigma=\frac{1}{\sqrt{-g}}(\sqrt{-g}g^{\gamma\delta}\varsigma_{,\gamma})_{,\delta}$,
$T^{(m)}=g^{\gamma\delta}T^{(m)}_{\gamma\delta}$ and
$\Theta=g^{\gamma\delta}\Theta_{\gamma\delta}$. We consider an
isotropic seed source whose energy-momentum tensor has the form
\begin{equation}\label{3}
T_{\gamma\delta}^{(m)}=\rho
u_{\gamma}u_{\delta}+p(u_{\gamma}u_{\delta}-g_{\gamma\delta}),
\end{equation}
where $p$, $u_{\gamma}$ and $\rho$ represent pressure, four-velocity
and energy density, respectively. The influence of massive scalar
field on the physical aspects of static sphere are governed by the
following energy-momentum tensor
\begin{equation}\label{4}
T_{\gamma\delta}^\varsigma=\varsigma_{,\gamma;\delta}-g_{\gamma\delta}\Box\varsigma+\frac{\omega_{BD}}{\varsigma}
(\varsigma_{,\gamma}\varsigma_{,\delta}
-\frac{g_{\gamma\delta}\varsigma_{,\alpha}\varsigma^{,\alpha}}{2})-\frac{V(\varsigma)g_{\gamma\delta}}{2}.
\end{equation}

The anisotropic extension is obtained by describing the internal
geometry of the static sphere as
\begin{equation}\label{5}
ds^2=e^{\phi(r)}dt^2-(e^{\psi(r)}dr^2+r^2d\theta^2+r^2\sin^2\theta
d\varphi^2).
\end{equation}
The field equations, formulated in the presence of the anisotropic
source, are expressed as
\begin{eqnarray}\label{6}
e^{-\psi}\left(-\frac{1}{r^2}+\frac{\psi'}{r}\right)+\frac{1}{r^2}&=&
\frac{1}{\varsigma}(\rho+\sigma\Theta_0^0+T_0^{0\varsigma}),\\\label{7}
e^{-\psi}\left(\frac{1}{r^2}+\frac{\phi'}{r}\right)-\frac{1}{r^2}&=&
\frac{1}{\varsigma}(p-\sigma\Theta_1^1-T_1^{1\varsigma}),\\\label{8}
\frac{e^{-\psi}}{4}\left(2\phi''+\phi'^2-\psi'\phi'+2\frac{\phi'-\psi'}{r}\right)&=&
\frac{1}{\varsigma}(p-\sigma\Theta_2^2-T_2^{2\varsigma}),
\end{eqnarray}
where
\begin{eqnarray}\nonumber
T_0^{0\varsigma}&=&e^{-\psi}\left[\varsigma''+\left(\frac{2}{r}-\frac{\psi'}{2}
\right)\varsigma'+\frac{\omega_{BD}}{2\varsigma}\varsigma'^2-e^\psi\frac{V(\varsigma)}
{2}\right],\\\nonumber
T_1^{1\varsigma}&=&e^{-\psi}\left[\left(\frac{2}{r}+\frac{\phi'}
{2}\right)\varsigma'-\frac{\omega_{BD}}{2\varsigma}\varsigma'^2-e^\psi\frac{V(\varsigma)}{2})\right],\\\nonumber
T_2^{2\varsigma}&=&e^{-\psi}\left[\varsigma''+\left(\frac{1}{r}-\frac{\psi'}
{2}+\frac{\phi'}{2}\right)\varsigma'+\frac{\omega_{BD}}{2\varsigma}\varsigma'^2-e^\psi\frac{V(\varsigma)}{2}
\right].
\end{eqnarray}
Here differentiation with respect to the radial coordinate is
denoted by $'$. Moreover, the wave equation (\ref{2}), containing
information about the evolution of the massive scalar field, takes
the form
\begin{eqnarray}\nonumber
\Box\varsigma&=&-e^{-\psi}\left[\left(\frac{2}{r}-\frac{\psi'} {2}
+\frac{\phi'}{2}\right)\varsigma'+\varsigma''\right]\\\label{2*}
&=&\frac{1}{2\omega_{BD}+3}\left[T^{(m)}+\Theta+
\left(\varsigma\frac{dV(\varsigma)}{d\varsigma}-2V(\varsigma)\right)\right].
\end{eqnarray}
The system of field equations (\ref{6})-(\ref{8}) correspond to an
anisotropic fluid distribution for $\Theta_1^1\neq\Theta_2^2$.

\section{Gravitational Decoupling}

Currently, the system (\ref{6})-(\ref{8}) involves unknown metric
potentials, state determinants, massive scalar field and anisotropic
source. Thus, we require additional constraints to reduce the
degrees of freedom and close the system. In this regard, the MGD
technique is an efficient method for decoupling the complex field
equations into simpler sets by transforming the radial metric
component as
\begin{eqnarray}\label{11}
e^{-\psi(r)}\mapsto\lambda(r)+\sigma \nu(r),
\end{eqnarray}
where the deformation function ($\nu(r)$) encodes the translation in
the radial metric function. The temporal metric component is not
transformed and remains unchanged. It must be noted that the linear
mapping does not affect the spherical geometry of the structure.
Moreover, we have used the parameter $\sigma$ in Eq.(\ref{11}) to
guarantee that the deformation in the radial metric component
vanishes for $\sigma=0$ leaving behind the unchanged
$(\phi,~\lambda)$-frame corresponding to the isotropic sector.
Plugging the transformation in Eqs.(\ref{6})-(\ref{8}) and setting
$\sigma=0$ yields the first set as
\begin{eqnarray}\nonumber
\rho&=&-\frac{1}{2r^2\varsigma(r)}[r^2\omega\lambda(r)\varsigma'^2(r)-r^2\varsigma(r)V(\varsigma)
+r\varsigma(r)(r\lambda'(r)\varsigma'(r)+2r\lambda(r)\varsigma''(r)\\\label{12}
&+&4\lambda(r)\varsigma'(r))+2\varsigma^2(r)
(r\lambda'(r)+\lambda(r)-1)],\\\nonumber
p&=&\frac{1}{r^2}[\varsigma(r)(r\lambda(r)
\phi'(r)+\lambda(r)-1)]+\frac{1}{2r\varsigma(r)}[\lambda(r)\varsigma'(r)(\varsigma(r)(r\phi'(r)+4)\\\label{13}
&-&r\omega_{BD}\varsigma'(r))]-\frac{V(\varsigma)}{2},\\\nonumber
p&=&\frac{1}{4r\varsigma(r)}[\varsigma(r)\lambda'(r)(\varsigma(r)(r\phi'(r)+2)
+2r\varsigma'(r))+\lambda(r)(2\varsigma(r)\varsigma'(r)\\\nonumber
&\times&((r\phi'(r)+2)+2r\varsigma''(r))+\varsigma^2(r)(2r\phi''(r)+r\phi'^2(r)+2\phi'(r))\\\label{14}
&+&2r\omega_{BD}\varsigma'^2(r))-2r\varsigma(r)V(\varsigma)].
\end{eqnarray}
The above system corresponds to the perfect fluid configuration
exclusively as the anisotropic effects are excluded. The isotropic
source observes the following conservation equation
\begin{equation}\label{14*}
T^{1'(\text{eff})}_{1}-\frac{\phi'(r)}{2}
(T^{0(\text{eff})}_{0}-T^{1(\text{eff})}_{1})=0.
\end{equation}
Moreover, for $\sigma=0$, the massive scalar field is related to the
co-ordinate frame $(\phi, \lambda)$ as
\begin{eqnarray*}
&&\frac{1}{r\varsigma(r)}\bigg(\varsigma^2
\left(\lambda(r)\left(2r^2\phi''(r)+r^2\phi'^2-2r
\phi'(r)-4\right)+r\left(r\phi'(r)+2\right)\lambda'(r)+4\right)\\
&&+4r^2
\omega_{BD}\lambda(r)\varsigma'^2+2r\varsigma(r)\left(r\lambda'(r)\varsigma'(r)-2
\lambda(r)\left(\varsigma'(r)-r\varsigma''(r)\right)\right)\bigg)=0.
\end{eqnarray*}
The second set, encompassing the terms related to the additional
source, is expressed as
\begin{eqnarray}\nonumber
\Theta_0^0&=&\frac{-1}{2r^2\varsigma(r)}[r\varsigma(r)\nu'(r)(r\varsigma'(r)+2\varsigma(r))+\nu(r)
(r^2\omega_{BD}\varsigma'^2(r)+2r\varsigma(r)\\\label{15}
&\times&(r\varsigma''(r)+2\varsigma'(r))+2\varsigma^2(r))],\\\nonumber
\Theta_1^1&=&\frac{-1}{2r^2\varsigma(r)}[\nu(r)(-r^2\omega_{BD}\varsigma'(r)^2+r\varsigma(r)(r
\phi'(r)+4)\varsigma'(r)+2\varsigma^2(r)\\\label{16}
&\times&(r\phi'(r)+1))],\\\nonumber
\Theta_2^2&=&\frac{-1}{4\varsigma(r)}[2\varsigma(r)(r
\nu'(r)\varsigma'(r)+\nu(r)((r\phi'(r)+2)\varsigma'(r)+2r\varsigma''(r)))\\\nonumber
&+&\varsigma^2(r)(\nu'(r)(r\phi'(r)+2)+\nu(r)(2r\phi''(r)+r\phi'^2(r)+2\phi'(r)))\\\label{17}
&+&2r\omega_{BD}\nu(r)\varsigma'^2(r)].
\end{eqnarray}
The energy and momentum of the anisotropic source is conserved as
\begin{equation}\label{17*}
\Theta^{1'(\text{eff})}_{1}-\frac{\phi'(r)}{2}(\Theta^{0(\text{eff})}_{0}-\Theta^{1(\text{eff})}_{1})
+\frac{2}{r}(\Theta^{2(\text{eff})}_{2}-\Theta^{1(\text{eff})}_{1})=0,
\end{equation}
where
\begin{eqnarray*}
\Theta^{0(\text{eff})}_0&=&\frac{1}{\varsigma}\left(\Theta^0_0+\frac{1}{2}
\nu'(r)\varsigma'(r)+\nu(r)\varsigma''(r)+\frac{\omega_{BD}\nu(r)\varsigma'^2(r)}{2\varsigma(r)}
+\frac{2\nu(r)\varsigma '(r)}{r}\right),\\
\Theta^{1(\text{eff})}_1&=&\frac{1}{\varsigma}\left(\Theta^1_1+\frac{1}{2}
\nu(r)\phi'(r)\varsigma'(r)-\frac{\omega_{BD}\nu(r)\varsigma'^2(r)}{2\varsigma(r)}+\frac{2
\nu(r)\varsigma'(r)}{r}\right),\\
\Theta^{2(\text{eff})}_2&=&\frac{1}{\varsigma}\left(\Theta^2_2+\frac{1}{2}
\nu'(r)\varsigma'(r)+\frac{1}{2}\nu(r)\phi'(r)\varsigma'(r)+\nu(r)\varsigma''(r)\right.\\
&+&\left.\frac{\omega_{BD}\nu(r)\varsigma'^2(r)}{2\varsigma(r)}+\frac{\nu(r)\varsigma'(r)}{r}\right).
\end{eqnarray*}
Thus, Tolman-Oppenheimer-Volkoff (TOV) equation corresponding to the entire
system is obtained by combining Eqs.(\ref{14*}) and (\ref{17*}) as
\begin{equation}\label{500}
T^{1'(\text{eff})}_{1}+\Theta^{1'(\text{eff})}_{1}-\frac{\phi'(r)}{2}
(T^{0(\text{eff})}_{0}-T^{1(\text{eff})}_{1}+\Theta^{0(\text{eff})}_{0}-\Theta^{1(\text{eff})}_{1})
+\frac{2}{r}(\Theta^{2(\text{eff})}_{2}-\Theta^{1(\text{eff})}_{1})=0.
\end{equation}

The MGD scheme restricts the exchange of matter or energy between
the two sources (seed and additional) by conserving them
individually. As a consequence of disintegrating the field equations
via MGD, the two systems can be solved separately. A well-behaved
isotropic spacetime can specify the system (\ref{12})-(\ref{14})
which minimizes the degrees of freedom as the second system contains
the deformation function and components of $\Theta_{\gamma\delta}$
as undetermined variables only. In the next section, we attain a
solution for the anisotropic setup by assuming a well-known ansatz
along with an additional constraint.

\section{Anisotropic Solutions}

\subsection{Solution I}

Tolman V ansatz (one of the eight solutions proposed by Tolman)
describes a spherical structure with infinite density and pressure
at the center \cite{30}. Zubair and Azmat \cite{31a} evaluated the
anisotropic counterpart of Tolman V via the MGD method. Recently,
Jasim et al. \cite{31b} employed this spherical solution to study
charged strange stars. Tolman V spacetime is defined as
\begin{equation}
ds^2=r^{2n}B^2dt^2-\bigg(\left(\frac{-1-2n+n^2}{\left(\frac{r}{F}\right)^W(1+2n-n^2)-1}\right)
dr^2+r^2d\theta^2+r^2\sin^2\theta d\varphi^2\bigg),\label{18}
\end{equation}
where $F,~B$ and $n$ are unknown constants with
$W=\frac{2(1+2n-n^2)}{1+n}$. We set
$\lambda=\frac{1+2n-n^2}{1-(1+2n-n^2)\left(\frac{r}{F}\right)^W}$
and determine the isotropic sector via the above spacetime. For this
purpose, we evaluate the unknown parameters through the continuity
of first and second fundamental forms at the junction ($\Sigma$) of
exterior and interior geometries. The exterior of an uncharged
static astrophysical object is defined via Schwarzschild metric
given as
\begin{equation}\label{20*}
ds^2=\frac{1}{r}(r-2M)dt^2-\bigg(\frac{1}{\frac{1}{r}(r-2M)}dr^2
+r^2d\theta^2+r^2\sin^2\theta d\varphi^2\bigg),
\end{equation}
where $M$ is the mass of the structure. The boundary conditions at
$r=R$ ($R$ is the radius of compact object)
\begin{eqnarray*}
(g_{\gamma\delta}^-)_{\Sigma}&=&(g_{\gamma\delta}^+)_{\Sigma},\quad
(p_{r})_{\Sigma}=0,\\
(\varsigma^-(r))_\Sigma&=&(\varsigma^+(r))_\Sigma,\quad
(\varsigma'^-(r))_\Sigma=(\varsigma'^+(r))_\Sigma.
\end{eqnarray*}
are utilized to determine the constants as
\begin{eqnarray}\label{20}
B&=&\frac{\sqrt{M} R^{\frac{R}{4 M-2 R}}}{\sqrt{-\frac{M}{2
M-R}}},\\\label{21} n&=&\frac{M}{R-2 M},\\\label{22} W&=&-\frac{2
\left(M^2+2 M R-R^2\right)}{2 M^2-3 M R+R^2},\\\label{23} F&=&R
\left(\frac{(R-2 M)M^2}{\left(-M^2-2 M
R+R^2\right)R}\right)^{-\frac{1}{W}}\\\nonumber
\omega_{BD}&=&\frac{2 M-R}{8M^2((n-2)n-1)}\left(-((n-2)n-1)(2M-R)
\left(\frac{R}{F}\right)^W(m_{\varsigma}R\right.\\\nonumber
&\times&\left.(2M-R)-4)-4m_{\varsigma}
M^2R+4m_{\varsigma}MR^2-m_{\varsigma}R^3+8Mn^3\right.\\\label{24}
&-&\left.24M n^2+8Mn+16M-8n^3 R+12n^2R+16nR\right).
\end{eqnarray}
Here, the corresponding scalar field has been calculated by
following the technique in \cite{24}.

The governing parameters of the anisotropic counterpart of Tolman V
are expressed as
\begin{eqnarray}\nonumber
\rho&=&\frac{-1}{2r^2\varsigma}\left(r\varsigma\left(r
\varsigma'\left(\sigma \nu'(r)+\frac{\zeta_1}{r}\right)+2\sigma
\nu(r)\left(r\varsigma''+2\varsigma'\right)+2\zeta_2\left(r\varsigma''\right.\right.\right.\\\nonumber
&+&\left.\left.\left.2\varsigma'\right)\right)+2\varsigma^2
\left(\sigma r \nu'(r)+\sigma
\nu(r)+\zeta_1+\zeta_2-1\right)+r^2\omega_{BD}\varsigma'^2
\left(\sigma \nu(r)\right.\right.\\\label{28}
&+&\left.\left.\zeta_2\right)-r^2\varsigma
V(\varsigma)\right),\\\nonumber p_r&=&\frac{\varsigma}{r^2}
\left(\nu(r) (\sigma +2 \sigma  n)+\zeta_2(2n+1)
-1\right)+\frac{\varsigma ' \left(2 (n+2) \varsigma-r \omega_{BD}
\varsigma'\right)}{2 r\varsigma}\\\label{29} &\times& \left(\sigma
\nu(r)+\zeta_2\right)-\frac{V(\varsigma)}{2},\\\nonumber
p_\perp&=&\frac{\varsigma}{2r^2}\left(\sigma(n+1)r \nu'(r)+2\sigma
n^2\nu(r)+\frac{1}{\zeta_3^2}((n-2)n-1)\left(2n-n^2+1)\right.\right.\\\nonumber
&\times&\left.\left.\left(2n^2-(n+1)W\right)
\left(\frac{r}{F}\right)^W-2n^2\right)\right)+\left(\sigma
\nu(r)+\zeta_2\right)\left(\frac{\varsigma'}{2}\right.\\\label{30}
&\times&\left.\left(\frac{\sigma \nu'(r)+\frac{\zeta_1}{r}}{\sigma
\nu(r)+\zeta_2}+\frac{2n}{r}+\frac{2}{r}\right)+\varsigma''
+\frac{\omega_{BD}\varsigma'^2}{2\varsigma}\right)-\frac{V(\varsigma)}{2},
\end{eqnarray}
where $\zeta_1=\frac{\left(n^2-2 n-1\right)^2 W
\left(\frac{r}{F}\right)^W}{\zeta_3}^2,~\zeta_2=\frac{-n^2+2
n+1}{\zeta_3}$ and $\zeta_3=\left(\left(n^2-2 n-1\right)
\left(\frac{r}{F}\right)^W\right.$ $\left.+1\right)$ with anisotropy
$\Delta=p_\perp-p_r$.
The matter variables are completely specified
if the deformation function is known. For this purpose, a constraint
on the deformation function or appropriate EoS is applied to the
anisotropic structure.

In this work, we develop an anisotropic model for strange quark
stars by implementing the MIT bag model. It is hypothesized that the
core of a quark star is composed of three flavors of quarks: up
$(u)$, strange $(s)$, and down $(d)$. We assume that the quarks are
massless and non-interacting in nature. According to the bag model,
the pressure and density of quark matter read
\begin{eqnarray}\label{7'}
p_r=\sum_{f}p^f-\mathcal{B},\quad \rho=\sum_{f}\rho^f+\mathcal{B},
\quad f=u,~d,~s,
\end{eqnarray}
where $p^f$ and $\rho^f$ correspond to the pressure and density of
each flavor, respectively. Kapusta \cite{31'} formulated the EoS for
strange quark matter through the relation $\rho^f=3p^f$ as
\begin{equation}\label{9}
3p_r=\rho-4\mathcal{B}.
\end{equation}
Utilizing Eqs.(\ref{28}) and (\ref{29}) in the bag model, we obtain
the following differential equation
\begin{eqnarray}\nonumber
&&\left(\sigma r \nu'(r)\varsigma'+\sigma \nu(r)\left(2(3
n+8)\varsigma'+2
r\varsigma''\right)+2\zeta_2r\varsigma''+\zeta_1\varsigma'+6n\zeta_2
\varsigma'\right.\\\nonumber &&+\left.16 \zeta_2\varsigma'-4 r
V(\varsigma)+8r\mathcal{B}\right)+\frac{2\varsigma}{r} \left(\sigma
r \nu'(r)+\sigma(6n+4)\nu(r)+\zeta_1\right.\\\label{27}
&&\left.+(6n+4)\zeta_2-4\right)-2r^2\omega_{BD}\varsigma'^2
\left(\sigma \nu(r)+\zeta_2\right)=0,
\end{eqnarray}
which is solved along with the wave equation using numerical
technique for central conditions
($\varsigma(0)=0.2,~\varsigma'(0)=0$ and $\nu(0)=\nu_c$) with
$V(\varsigma)=\frac{1}{2}m_{\varsigma}^2\varsigma^2$.

In SBD gravity, the values of coupling parameter greater than
$\frac{-3}{2}$ are allowed for $m_{\varsigma}> 10^{-4}$ (in
dimensionless units). Therefore, the wave equation is solved
numerically for $m_{\varsigma}=0.001$. Yazadjiev et al. \cite{55a}
have studied the structure of slowly rotating neutron stars by
considering $m_\varsigma=0.001$. Furthermore, the constants $B,F,n$
remain unchanged for the anisotropic sphere whereas $\omega_{BD}$ is
an undetermined free parameter. Thus, we choose the value of
$\omega_{BD}$ as given in Eq.(\ref{24}). The constants are computed
by utilizing the observed mass and radii of three quark star
candidates (PSR J1614-2230, Her X1 and 4U 1608-52) for $\sigma=
0.2,~0.9$ and $\mathcal{B}=60MeV/fm^3$. The values of $\omega_{BD}$
and $\nu_c$ for the considered stars are given in Table \textbf{1}.
It is noted that $\omega_{BD}$ attains values greater than
$\frac{-3}{2}$ for all stellar candidates. The physical properties
of the anisotropic stellar model are examined in the next section.
\begin{table} \caption{Values of different parameters corresponding to
solution I for $m_\varsigma=0.001$ and $\mathcal{B}=60MeV/fm^3$.}
\begin{center}
\begin{tabular}{|c|c|c|c|}
\hline &PSR J1614-2230 & Her X-1& 4U 1608-52  \\
\hline $M~(M_{\bigodot})$ &1.97 & 0.88 & 1.74 \\
\hline $R~(km)$ &13 & $7.7$ & 9.3\\
\hline $\omega_{BD}$& 9.750 & 15.353 & 6.593\\
\hline $\nu_c~(\sigma=0.2)$ & -5.1 & -3.6 & -6.8\\
\hline $\nu_c~(\sigma=0.9)$  & -1.132 & -0.8 & -1.5\\ \hline
\end{tabular}
\end{center}
\end{table}

\subsubsection{Physical Features of Anisotropic Model}

The metric potentials of a well-behaved solution must be positive
and increase monotonically so that pressure and density follow a
decreasing trend away from the center \cite{134}-\cite{134b}.
Figures \textbf{1} and \textbf{2} show that the deformed radial
metric obtained via decoupling is free from singularity for both
values of $\sigma$. Moreover, as most of the matter is concentrated
at the core of the star therefore, the state parameters must observe
a monotonically decreasing behavior away from the center. The
graphical representations of Eqs.(\ref{28})-(\ref{30}) in Figures
\textbf{3} and \textbf{4} reveal that the strange star model has
maximum density and radial/tangential pressures in the core.
Furthermore, radial pressure is zero at the surface for the
considered stellar objects which is necessary for the equilibrium of
the compact object \cite{135, 135a}. It is observed from the plots
of anisotropy that $p_\perp>p_r$ for 4U 1608-52 whereas anisotropy
related to the remaining stars is negative, i.e., $p_\perp<p_r$. It
is worthwhile to mention here that the GR analog exhibits similar
behavior \cite{31a}.
\begin{figure}\center
\epsfig{file=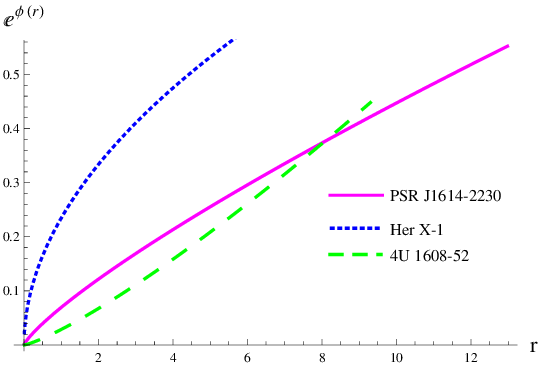,width=0.4\linewidth}\epsfig{file=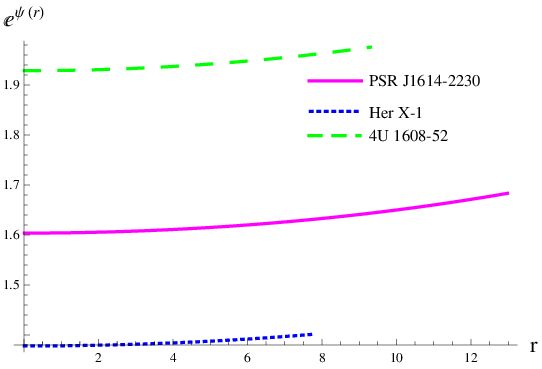,width=0.4\linewidth}
\caption{Behavior of $e^\phi$ and $e^\psi$ for anisotropic Tolman V
solution for $\sigma=0.2$.}
\end{figure}
\begin{figure}\center
\epsfig{file=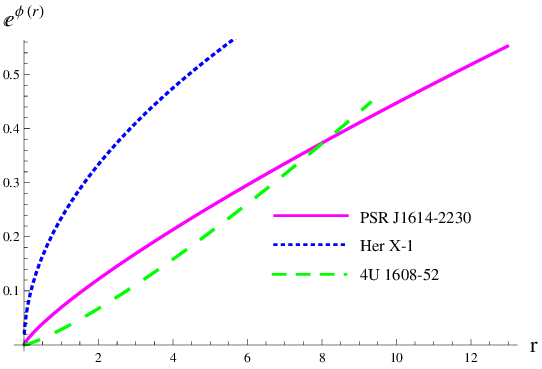,width=0.4\linewidth}\epsfig{file=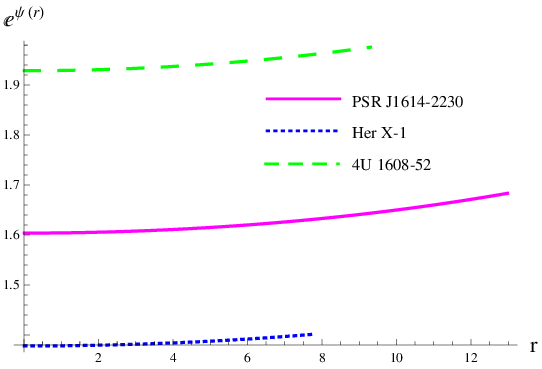,width=0.4\linewidth}
\caption{Behavior of $e^\phi$ and $e^\psi$ for anisotropic Tolman V
solution for $\sigma=0.9$.}
\end{figure}
\begin{figure}\center
\epsfig{file=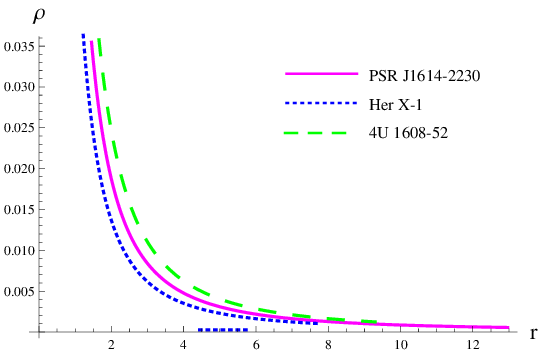,width=0.4\linewidth}\epsfig{file=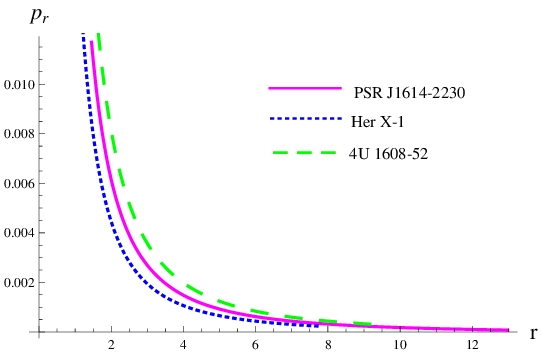,width=0.4\linewidth}
\epsfig{file=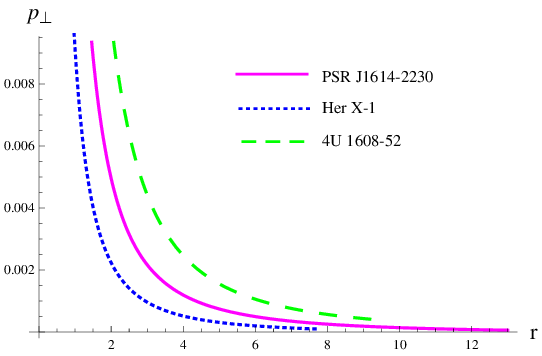,width=0.4\linewidth}\epsfig{file=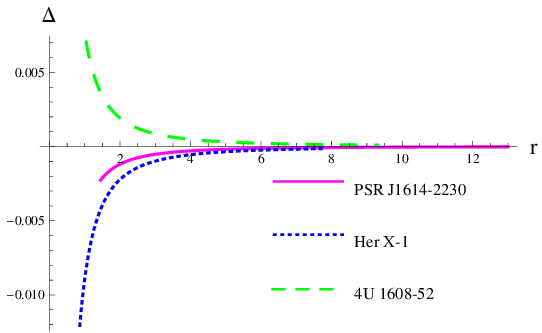,width=0.4\linewidth}
\caption{Behavior of $\rho,~p_r,~p_\perp$ (in $km^{-2}$) and
$\Delta$ of anisotropic Tolman V solution for $\sigma=0.2$.}
\end{figure}
\begin{figure}\center
\epsfig{file=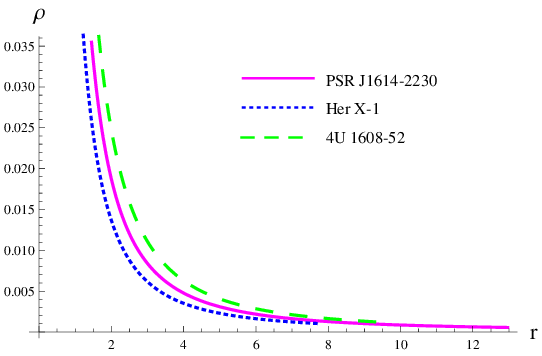,width=0.4\linewidth}\epsfig{file=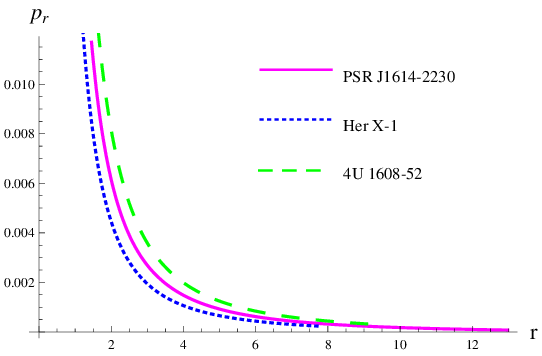,width=0.4\linewidth}
\epsfig{file=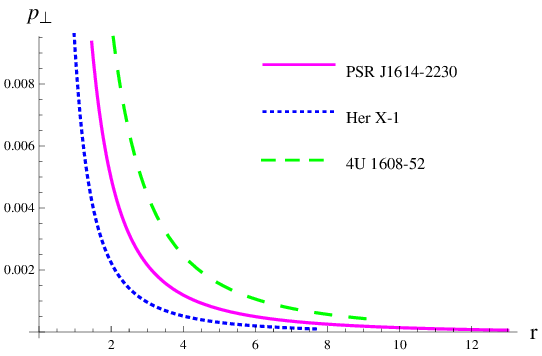,width=0.4\linewidth}\epsfig{file=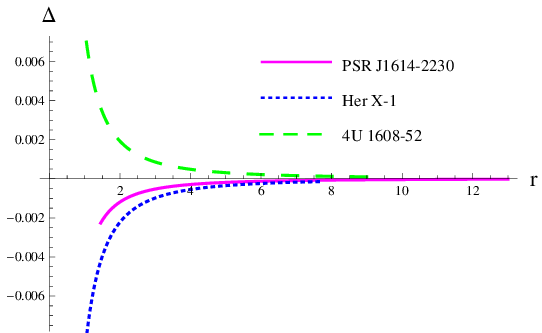,width=0.4\linewidth}
\caption{Behavior of $\rho,~p_r,~p_\perp$ (in $km^{-2}$) and
$\Delta$ of anisotropic Tolman V solution for $\sigma=0.9$.}
\end{figure}
\begin{figure}\center
\epsfig{file=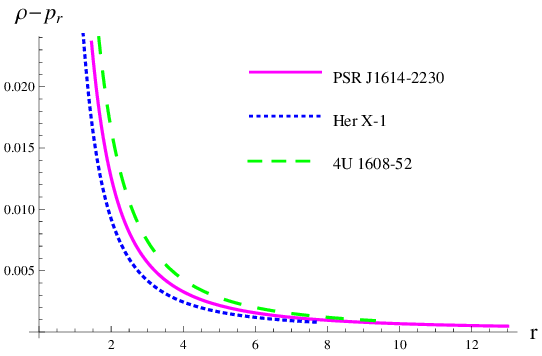,width=0.4\linewidth}
\epsfig{file=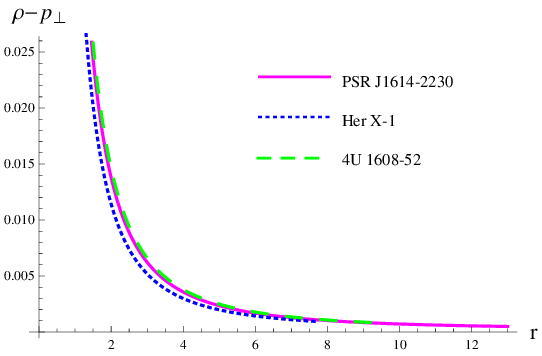,width=0.4\linewidth} \caption{DEC (in
$km^{-2}$) for extended Tolman V solution for $\sigma=0.2$.}
\end{figure}
\begin{figure}\center
\epsfig{file=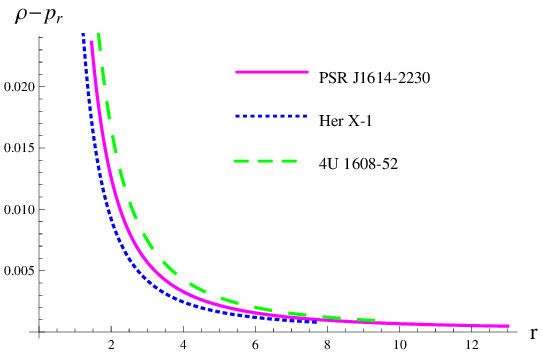,width=0.4\linewidth}\epsfig{file=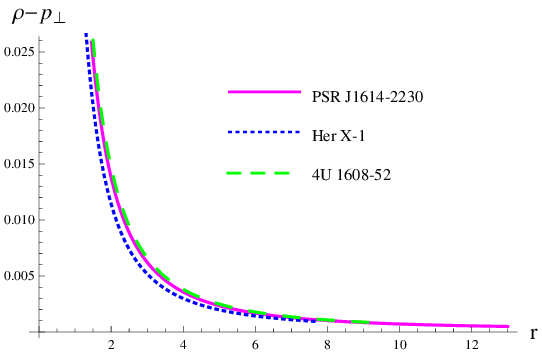,width=0.4\linewidth}
\caption{DEC (in $km^{-2}$) for extended Tolman V solution for $\sigma=0.9$.}
\end{figure}

As the interior of compact stellar structures consists of normal
matter, therefore the energy-momentum tensor describing the internal
configuration must obey the four energy bounds \cite{34}
\begin{eqnarray*}
&&\text{null energy condition:}\quad\rho+p_r\geq0,\quad\rho+p_\perp\geq0,\\
&&\text{weak energy condition:}\quad\rho\geq0,\quad\rho+p_r\geq0,\quad\rho+p_\perp\geq0,\\
&&\text{strong energy condition:}\quad\rho+p_r+2p_\perp\geq0,\\
&&\text{dominant energy condition:}\quad\rho-p_r\geq0,\quad
\rho-p_\perp\geq0.
\end{eqnarray*}
The first three bounds demand that pressure components and energy
density remain positive throughout the setup. As the matter
variables presented in Figures \textbf{3} and \textbf{4} fulfil this
criterion, the anisotropic Tolman V is consistent with null, weak
and strong energy constraints. Figures \textbf{5} and \textbf{6}
indicate that the decoupled model obeys the fourth condition as
well. Thus, the constructed model represents a viable quark star.

\begin{figure}\center
    \epsfig{file=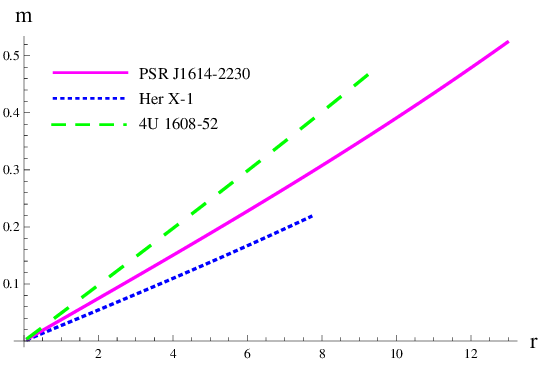,width=0.35\linewidth}\epsfig{file=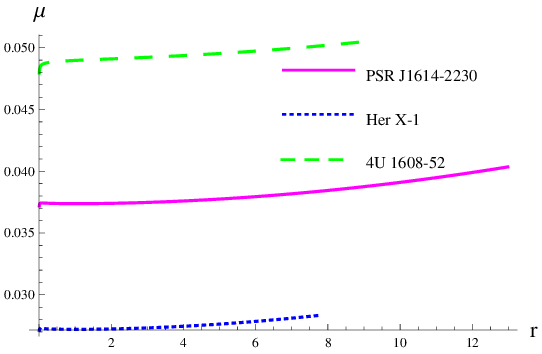,width=0.35\linewidth}
    \epsfig{file=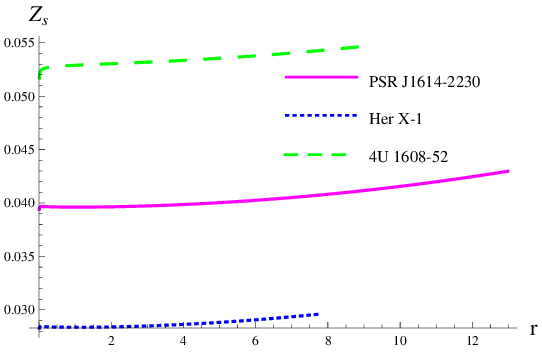,width=0.35\linewidth} \caption{Graphical
        representation  of  $m$ ($M_{\bigodot}$), $\mu$ and $Z_s$ corresponding to anisotropic
        Tolman V solution for $\sigma=0.2$.}
\end{figure}
The mass of the compact spherical structure is evaluated by solving
\begin{equation}\label{31}
m(r)=\frac{1}{2}\int_0^rr^2\rho dr.
\end{equation}
with $m(0)=0$. Buchdahl \cite{32} observed that the ratio of mass to
radius of compact objects (also known as compactness factor
$\mu(r)$) is less than 0.444. The compactness of the anisotropic
configuration lies below the required limit for the considered
values of the parameters (refer to Figures \textbf{7} and
\textbf{8}). Gravitational redshift is another interesting feature
of astrophysical objects. It measures the effect of gravitational
field of a massive object on light \cite{106} and is defined as
\cite{107}
\begin{equation*}
Z_s(r)=-1+\frac{1}{\sqrt{1-2u}}.
\end{equation*}
The value of the redshift function for cosmic objects composed of
perfect fluid cannot exceed 2. However, the upper limit for
anisotropic models changes to 5.211 \cite{33}. The range of redshift
parameter plotted in Figures \textbf{7} and \textbf{8} lies in the
admissible interval.
\begin{figure}\center
\epsfig{file=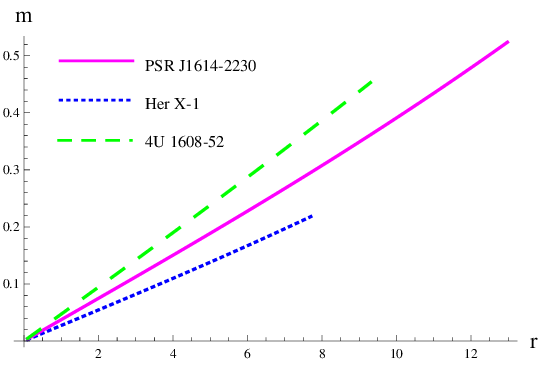,width=0.35\linewidth}\epsfig{file=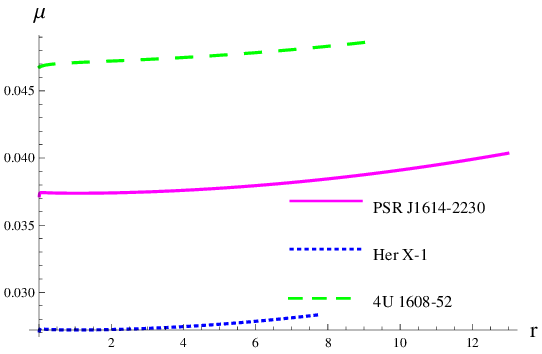,width=0.35\linewidth}
\epsfig{file=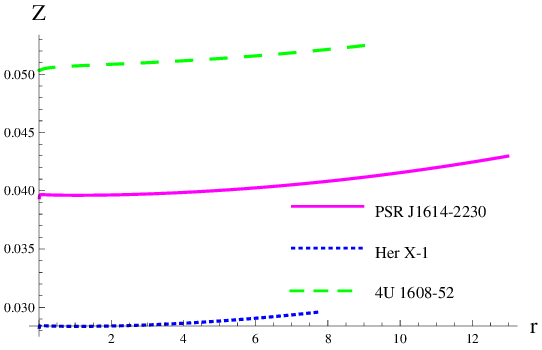,width=0.35\linewidth} \caption{Graphical
representation of $m$ ($M_{\bigodot}$), $\mu$ and $Z_s$ corresponding to anisotropic
Tolman V solution for $\sigma=0.9$.}
\end{figure}
\begin{figure}\center
    \epsfig{file=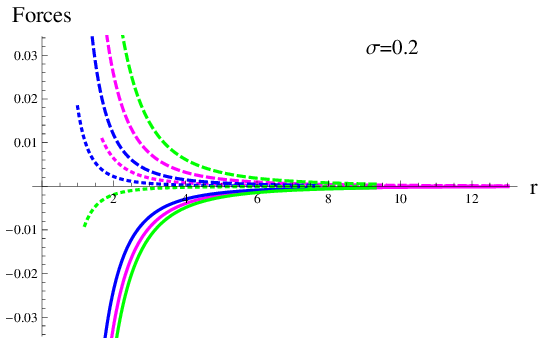,width=0.4\linewidth}\epsfig{file=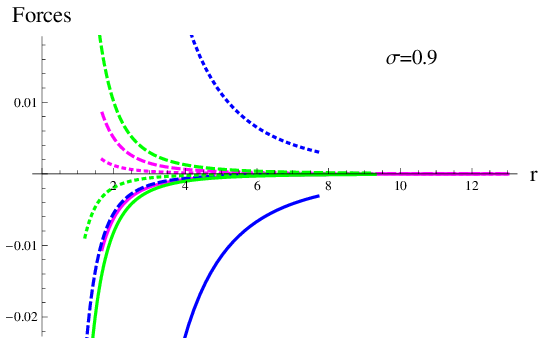,width=0.4\linewidth}
    \caption{Graphical representation of $f_h$ (solid), $f_g$ (dashed)
    and $f_a$ (dotted) for PSR J1614-2230 (pink), Her X-1 (blue) and 4U 1608-52 (green) corresponding to extended Tolman V
        solution with $\sigma=0.2$ and $0.9$.}
\end{figure}

The TOV equation (\ref{500}) is a significant tool to determine
if the anisotropic system is in a state of equilibrium.
The anisotropic static stellar structure maintains equilibrium if sum of the following forces equals zero
\begin{itemize}
\item gravitational force: $f_g=\frac{-\nu'}{2}(\rho+p_r)$,
\item hydrostatic force: $f_h=p'_r$,
\item anisotropic force: $f_a=\frac{2}{r}(p_r-p_\perp)$.
\end{itemize}
Figure \textbf{9} indicates that the three force balance
each other corresponding to the anisotropic version of Tolman V solution.
The extended version of Tolman V solution is stable if it complies
with the causality condition which states that velocity of light
always exceeds the rate at which sound travels, i.e.,
$1>v_\perp^2>0$ and $1>v_r^2>0$ ($v_\perp^2=\frac{dp_\perp}{d\rho}$
and $v_r=\frac{dp_r}{d\rho}$ represent transverse and radial
components of sound, respectively). It must be noted that the
propagation of sound waves changes with a change in the medium.
Figures \textbf{10} and \textbf{11} depict that the extended solution
is consistent with the causality criterion for higher as well as
lower values of decoupling parameter. In order to inspect the
anisotropic system for potential stability, we implement Herrera's
cracking criterion. According to this approach, if a disturbance in
the equilibrium of the system leads to radial forces of different
signs, the configuration cracks. The condition
$0<|v_\perp^2-v_r^2|<1$ ensures that no cracking appears within the
compact sphere. This condition holds true for the current setup as
shown in Figures \textbf{10} and \textbf{11}.
\begin{figure}\center
    \epsfig{file=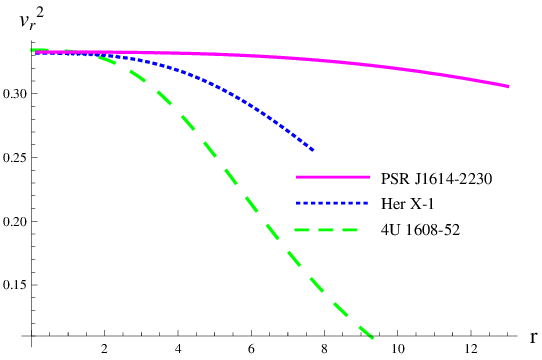,width=0.4\linewidth}\epsfig{file=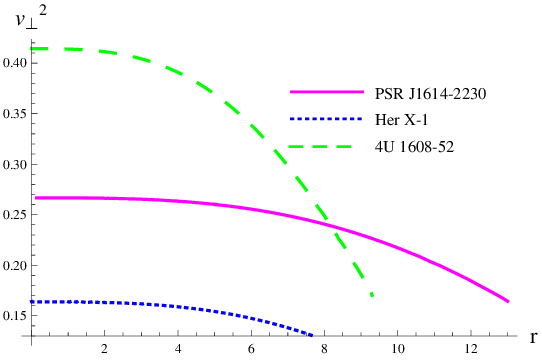,width=0.4\linewidth}
    \epsfig{file=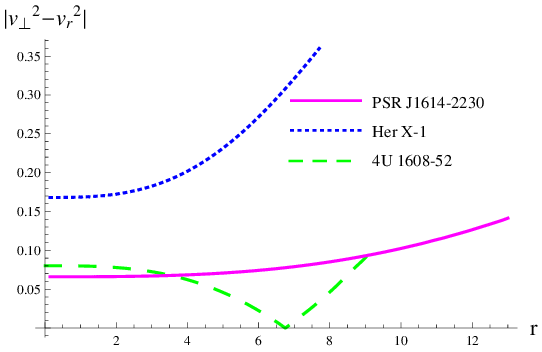,width=0.4\linewidth}\epsfig{file=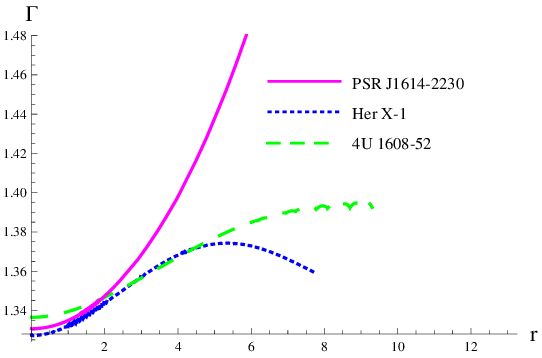,width=0.4\linewidth}
    \caption{Graphical representation of sound velocities,
        $|v_\perp^2-v_r^2|$ and adiabatic index of extended Tolman V
        solution for $\sigma=0.2$.}
\end{figure}
\begin{figure}\center
    \epsfig{file=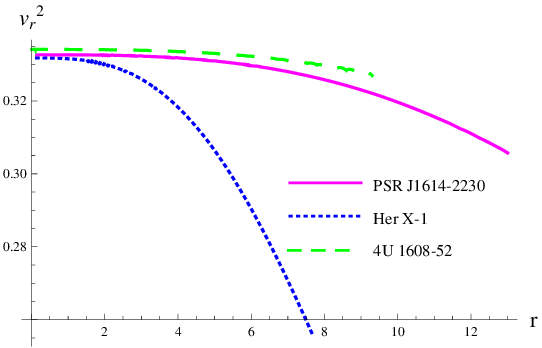,width=0.4\linewidth}\epsfig{file=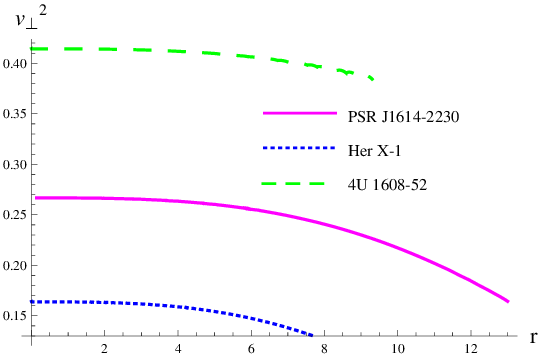,width=0.4\linewidth}
    \epsfig{file=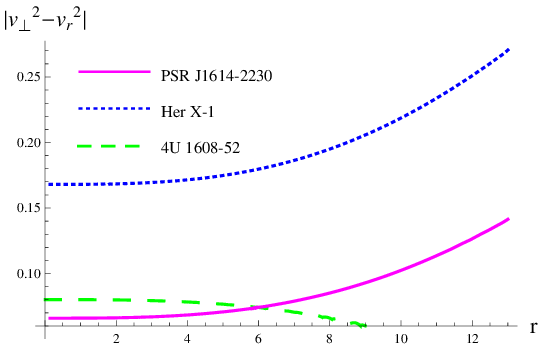,width=0.4\linewidth}\epsfig{file=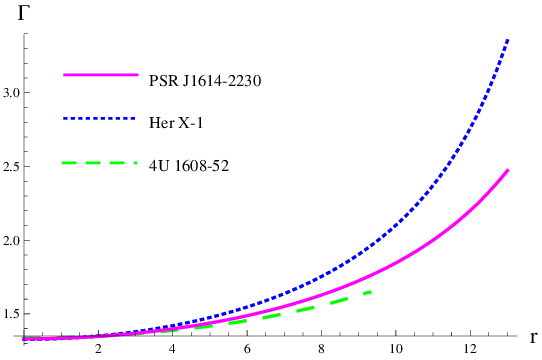,width=0.4\linewidth}
    \caption{Graphical representation  of sound velocities,
        $|v_\perp^2-v_r^2|$ and adiabatic index of extended Tolman V
        solution for $\sigma=0.9$.}
\end{figure}

Chandrasekhar \cite{200} investigated the dynamical stability of a spherical
structure in GR by using the limiting value of adiabatic index. Many authors
have used this definition to check the stability of various compact star
models \cite{201}. If a stellar model obeys a stiff EoS, it is compact and difficult to
compress as it generates large outward pressure in response to a
small increase in density. Adiabatic index measures the stiffness of
the EoS as
\begin{equation*}
\Gamma=\frac{dp_r}{d\rho}\frac{\rho+p_r}{p_r}
=v_r^2\frac{p_r+\rho}{p_r}.
\end{equation*}
The value of adiabatic index greater than $\frac{4}{3}$ corresponds
to a stiff matter distribution \cite{41}. The adiabatic index for
different values of $\sigma$ is displayed in Figures \textbf{10} and
\textbf{11}. The generated extension of Tolman V observes the
required restraint under the influence of massive scalar field.
It is worth mentioning that Harrison-Zeldovich-Novikov provided another
useful technique to investigate the stability of a non-rotating static sphere.
According to this method, the deformed mass must be an
increasing function of its central density $(\rho_0)$
under small radial perturbations, i.e., $\frac{\partial M}{\partial\rho_0}>0$.
In order to check the stability through this definition, it is necessary to
express the mass in terms of central density. However, in our work,
an explicit expression of mass cannot be obtained due to the presence
of massive scalar field. Thus, we cannot apply this stability criterion in our study.

\subsection{Solution II}

We develop the second anisotropic strange star model by assuming
that the radial metric component related to the isotropic spacetime
has the following form
\begin{equation}\label{32}
\lambda=\left(\frac{\cosh \left(2 \left(a r^2+b\right)\right)+2 c
r^2+1}{\cosh \left(2 \left(a r^2+b\right)\right)+1}\right)^{-1},
\end{equation}
where $a,~b$ and $c$ are nonzero constants. This form of the metric
function increases monotonically towards the boundary and is free
from singularity. Thus, Eq.(\ref{32}) corresponds to a physically
valid metric potential. Moreover, Maurya and Maharaj \cite{130} have
obtained a physically acceptable solution by employing the ansatz in
Eq.(\ref{32}). We obtain the temporal metric component via
Karmarkar's embedding condition. A spherical spacetime belongs to
embedding class-two, i.e., it can be embedded in a six-dimensional
flat spacetime. Karmarkar \cite{131} obtained a condition through
the Gauss-Codazi equations which allows the embedding of a spherical
line element into a five-dimensional flat spacetime. To embed
$m$-dimensional space in an $(m+1)$-dimensional pseudo-Euclidean
space, the following condition must hold \cite{132}
\begin{equation}\nonumber
\mathbb{T}_{\gamma[\delta;\mu]}-\Gamma^{\lambda}_{\delta\mu}\mathbb{T}_{\gamma\lambda}
+\Gamma^\lambda_{\gamma[\delta}\mathbb{T}_{\mu]\lambda}=0\quad\text{and}\quad\mathcal{R}_{\gamma\delta\mu\upsilon}=2e\mathbb{T}_{\gamma[\mu}
\mathbb{T}_{\upsilon]\delta},
\end{equation}
where $\mathbb{T}_{\gamma\delta}$ and
$\mathcal{R}_{\gamma\delta\mu\upsilon}$ are the co-efficients of
second differential form and curvature tensor, respectively.
Furthermore, $e=\pm1$. Using this result, Karmarkar formulated a
necessary and sufficient embedding class-one condition as
\begin{equation}\nonumber
\mathcal{R}_{2323}\mathcal{R}_{0101}=\mathcal{R}_{1202}\mathcal{R}_{1303}+\mathcal{R}_{1212}\mathcal{R}_{0303},
\end{equation}
which for the ($\lambda$, $\phi$)-metric implies
\begin{equation}\label{300}
(\lambda (r)-1) \lambda (r) \left(2 \phi ''(r)+\phi
'(r)^2\right)-\lambda '(r) \phi '(r)=0.
\end{equation}
Different authors have employed Karmarkar condition to develop well-behaved stellar models \cite{203, 207}. Plugging Eq.(\ref{32}) in (\ref{300}) yields
\begin{equation}\label{33}
\phi=\ln\left(q_1^2 \left(\tan^{-1}\left(\sinh\left(a
r^2+b\right)\right)+q_2\right)^2\right),
\end{equation}
where $q_1=\alpha_1\frac{\sqrt{c}}{2a}$,
$q_2=\frac{2a\alpha_2}{\sqrt{c}\alpha_1}$, $\alpha_1$ and $\alpha_2$
are constants. The obtained temporal metric potential is an
increasing function of $r$ with no singularity.

The unknown constants, determined by matching the interior spherical
line element (corresponding to the metric components in
Eqs.(\ref{32}) and (\ref{33})) with the Schwarzschild spacetime,
read
\begin{eqnarray}\label{34}
c&=&-\frac{2 M \cosh ^2\left(a R^2+b\right)}{R^2(2
M-R)},\\\label{35}
q_1&=&\frac{\sqrt{R-2M}}{16\sqrt{\frac{a^2R^5
\left(2M^2-3MR+R^2\right)^2\ln\left(-\frac{R}{2 M-R}\right)^2
\text{sech}^2\left(a R^2+b\right)}{\left((R-2M)^2(m_\varsigma
R(2M-R)-4)-4\left(2M^2(\omega_{BD}+2)-R^2\right)\ln\left(-\frac{R}{2M-R}\right)\right)^2}}},\\\nonumber
q_2&=&\frac{-16aR^2(R-2M)(R-M)\ln\left(\frac{2M}{R-2M}+1\right)
\text{sech}\left(aR^2+b\right)}{(R-2M)^2(m_\varsigma R(2M-R)-4)+4
\left(R^2-2M^2(\omega_{BD}+2)\right)\ln\left(\frac{R}{R-2M}\right)}\\\label{36}
&-&\tan^{-1}\left(\sinh\left(aR^2+b\right)\right).
\end{eqnarray}
The field equations incorporating the anisotropic source related to
the considered setup take the form
\begin{eqnarray}\nonumber
\rho&=&\frac{-1}{2 r^2 \varsigma }\left(r \varsigma\left(r
\varsigma'\left(\frac{4 c r \left(2 a r^2 \sinh \left(2 \left(a
r^2+b\right)\right)-\cosh \left(2\left(a
r^2+b\right)\right)-1\right)}{(\zeta_5)^2}\right.\right.\right.\\\nonumber
&+&\left.\left.\left.\sigma \nu'(r)\right)+\frac{4 \cosh ^2\left(a
r^2+b\right) \left(r\varsigma ''(r)+2 \varsigma'\right)}{\zeta_5}+2
\sigma  \nu(r) \left(r \varsigma''+2 \varsigma
'\right)\right)\right.\\\nonumber
&+&\left.2\varsigma^2\left(r\sigma\nu'+\frac{8acr^3 \sinh
\left(2\left(ar^2+b\right)\right)}{\left(\zeta_5\right)^2}+\frac{r(3\cosh
\left(2 \left(a r^2+b\right)\right)+2c
r^2)}{\left(\zeta_5\right)^2}\right.\right.\\\label{37}
&+&\left.\left.\frac{3}{(\zeta_5)^2}+\sigma \nu\right)+r^2
\omega_{BD} \varsigma '^2 \left(\frac{2 \cosh^2\left(a
r^2+b\right)}{\zeta_5}+\sigma \nu\right)-r^2 \varsigma
V(\varsigma)\right),\\\nonumber
p_r&=&\frac{\varsigma'}{2r\varsigma}\left(\frac{2\cosh^2\left(ar^2+b\right)}{\zeta_5}+\sigma
\nu(r)\right) \left(\varsigma (r) \left(\frac{4 a
r^2\text{sech}\left(ar^2+b\right)}{\zeta_4}+4\right)\right.\\\nonumber
&-&\left.r\omega_{BD} \varsigma '(r)\right)+\frac{\varsigma (r)
\sigma\nu(r)\zeta_4\zeta_5\text{sech}\left(a r^2+b\right)
\left(\cosh \left(a  r^2+b\right)+4 a r^2\right)}{r^2
\zeta_4\zeta_5}\\\label{38}
&-&\varsigma\frac{2r^2\left(c\zeta_4-4a\cosh \left(a
r^2+b\right)\right)}{r^2\zeta_4\zeta_5}-\frac{V(\varsigma)}{2},\\\nonumber
p_\perp&=&\frac{\varsigma (r)}{2}  \left(\frac{2 \cosh ^2\left(a
r^2+b\right)}{\zeta_5}+\sigma\nu(r)\right) \left(\frac{8 a^2 r^2
\text{sech}^2\left(a r^2+b\right)}
{\left(\zeta_4\right)^2}\right.\\\nonumber
&+&\left.\frac{2ar\text{sech}\left(ar^2+b\right)\left(\frac{4 c r
\left(2 a r^2 \sinh \left(2\left(ar^2+b\right)\right)-\cosh
\left(2\left(ar^2+b\right)\right)-1\right)}
{\left(\zeta_5\right)^2}+\sigma  \nu'(r)\right)}{\zeta_4
\left(\frac{2\cosh ^2\left(ar^2+b\right)}{\zeta_5}+\sigma
\nu(r)\right)}\right.\\\nonumber
&+&\left.\frac{4ar\text{sech}\left(ar^2+b\right)}{r\zeta_4}
+\frac{\frac{4 c r \left(2 a r^2 \sinh \left(2 \left(a
r^2+b\right)\right)-\cosh\left(2\left(ar^2+b\right)\right)-1\right)}
{\left(\zeta_5\right)^2}+\sigma\nu'}{r\frac{2\cosh ^2\left(a
r^2+b\right)}{\zeta_5}+\sigma\nu(r)}\right)\\\nonumber
&+&\frac{4a\text{sech}^2\left(ar^2+b\right)\left(\cosh\left(a
r^2+b\right) \left(-\tan ^{-1}\left(\sinh
\left(ar^2+b\right)\right)-q_2\right)\right)}
{\left(\zeta_4\right)^2}\\\nonumber
&+&\left(\frac{2\cosh^2\left(ar^2+b\right)}{\zeta_5}+\sigma
\nu(r)\right)\left(\frac{\varsigma'(r)}{2}\left(\frac{4ar
\text{sech}\left(ar^2+b\right)}{\zeta_4}\right.\right.\\\nonumber
&+&\left.\left.\frac{\frac{4cr\left(2ar^2\sinh\left(2\left(a
r^2+b\right)\right)-\cosh\left(2\left(ar^2+b\right)\right)-1\right)}
{\left(\zeta_5\right)^2}+\sigma\nu'(r)}
{\frac{2\cosh^2\left(ar^2+b\right)}{\zeta_5}+\sigma\nu(r)}
+\frac{2}{r}\right)+\varsigma''(r)\right.\\\nonumber
&-&\left.\frac{4a\text{sech}^2\left(a r^2+b\right) \left(2 a
r^2\left(\sinh\left(ar^2+b\right)\zeta_4+1\right)\right)}
{\left(\zeta_4\right)^2}+\frac{\omega_{BD} \varsigma '^2}{2
\varsigma}\right)-\frac{V(\varsigma )}{2},\\\label{39}
\end{eqnarray}
where $\zeta_4=\tan^{-1}\left(\sinh \left(ar^2+b\right)\right)+q_2$
and $\zeta_5=\cosh\left(2\left(ar^2+b\right)\right)+2cr^2+1$.

Applying the MIT bag model to the spherical spacetime yields
\begin{eqnarray}\nonumber
&&\frac{1}{r \varsigma (r)}\left(2 \varsigma^2
\left(\frac{2\cosh^2\left(ar^2+b\right)}{\zeta_5}
\left(\frac{12ar^2\text{sech}\left(ar^2+b\right)} {\zeta_4}+4\right)
+\frac{\sigma\nu}{\zeta_4}\left(4\zeta_4
\right.\right.\right.\\\nonumber
&&\left.\left.+12ar^2\text{sech}\left(ar^2+b\right)\right)+\sigma
r\nu'-4+\frac{4cr^2}{\left(\zeta_5\right)^2}
\left(2ar^2\sinh\left(2\left(ar^2+b\right)\right)\right.\right.\\\nonumber
&&\left.\left.\left.-\cosh\left(2\left(ar^2+b\right)\right)-1\right)\right)+r\varsigma(r)
\left(\frac{24 a r^2 \varsigma '(r) \cosh \left(a
r^2+b\right)}{\zeta_4 \zeta_5}+\sigma\nu\right.\right.\\\nonumber
&&\left.\left.\times \left(\varsigma '(r) \left(\frac{12 a r^2
\text{sech}\left(a r^2+b\right)}{\zeta_4}+16\right)+2r
\varsigma''(r)\right)+\frac{4 r
\varsigma''\cosh^2\left(ar^2+b\right)}{\zeta_5}\right.\right.\\\nonumber
&&\left.\left.+\frac{32\varsigma'(r)\cosh^2\left(ar^2+b\right)}{\zeta_5}
+\frac{4cr^2\varsigma'(r)}{\left(\zeta_5\right)^2} \left(2 a r^2
\sinh \left(2 \left(a r^2+b\right)\right)+\sigma  r \nu' \varsigma
'(r)\right.\right.\right.\\\nonumber
&&\left.\left.\left.-\cosh\left(2 \left(a
r^2+b\right)\right)-1\right)-4 r V(\varsigma)+8 r
\mathcal{B}\right)-2 r^2 \omega_{BD}\varsigma'^2 \left(\sigma
\nu\right.\right.\\\label{40}
&&\left.\left.+\frac{2\cosh^2\left(ar^2+b\right)}{\zeta_5}\right)\right)=0.
\end{eqnarray}
The deformation function is computed through numerical solutions of
Eqs.(\ref{2*}) and (\ref{40}) for
$V(\varsigma)=\frac{1}{2}m_{\varsigma}^2\varsigma^2$,
$m_\varsigma=0.001,~\mathcal{B}=60MeV/fm^3$ and $\sigma=0.2,~0.9$
corresponding to the three stellar models employed in solution I.
Moreover, the central conditions of solution I are applied in this
scenario as well. The associated values of $\nu_c,~\omega_{BD},~a$
and $b$ are presented in Table \textbf{2}. Furthermore, the
constants $q_1$ and $q_2$ appear as free parameters in the
anisotropic distribution. Thus, we select the respective values of
these constants as given in Eqs.(\ref{35}) and (\ref{36}).

\subsubsection{Physical Features of Anisotropic Model}

\begin{table} \caption{Values of different parameters corresponding to solution II
        $m_\varsigma=0.001$ and $\mathcal{B}=60MeV/fm^3$.}
    \begin{center}
        \begin{tabular}{|c|c|c|c|}
            \hline &PSR J1614-2230 & Her X-1& 4U 1608-52  \\
            \hline $a$ &0.004 & 0.0026 & 0.006 \\
            \hline $b$ &0.85 & $0.85$ & 0.85\\
            \hline $\omega_{BD}$& 10 & 5 & 8\\
            \hline $\nu_c~(\sigma=0.2)$ & 5 & 5 & 5\\
            \hline $\nu_c~(\sigma=0.9)$  & 1.1 & 1.1 & 1.1\\ \hline
        \end{tabular}
    \end{center}
\end{table}
\begin{figure}\center
\epsfig{file=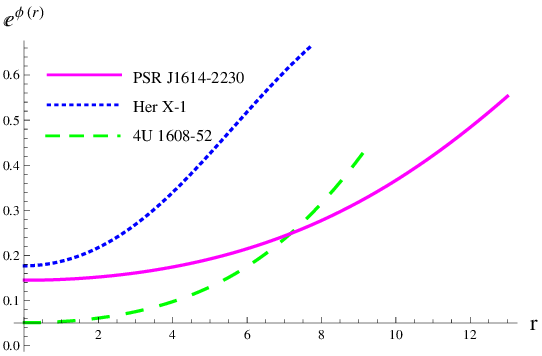,width=0.4\linewidth}\epsfig{file=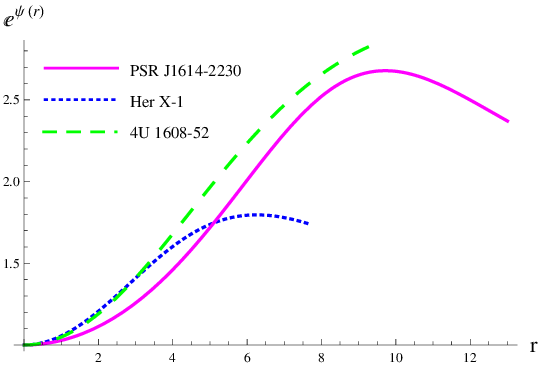,width=0.4\linewidth}
\caption{Temporal and radial metric components of anisotropic
solution II for $\sigma=0.2$.}
\end{figure}
\begin{figure}\center
    \epsfig{file=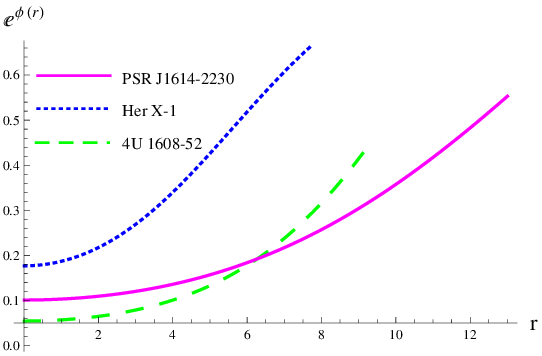,width=0.4\linewidth}\epsfig{file=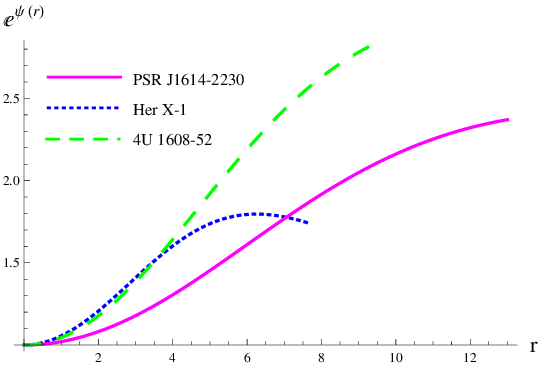,width=0.4\linewidth}
    \caption{Temporal and radial metric components of anisotropic
        solution II for $\sigma=0.9$.}
\end{figure}
\begin{figure}\center
    \epsfig{file=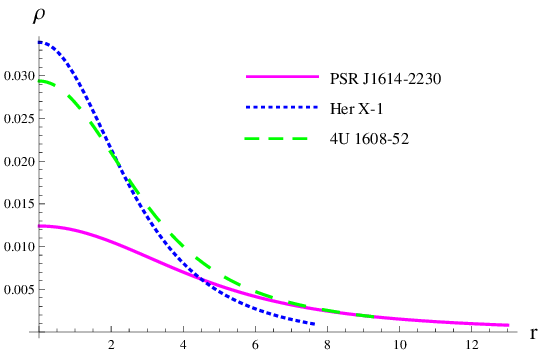,width=0.4\linewidth}\epsfig{file=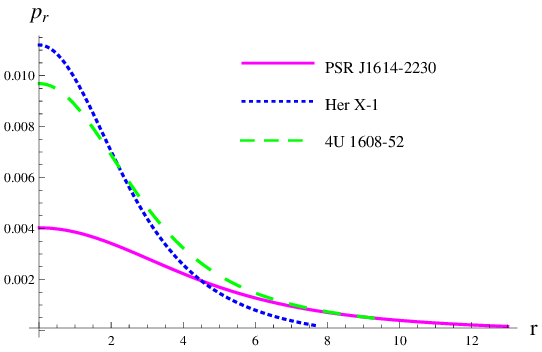,width=0.4\linewidth}
    \epsfig{file=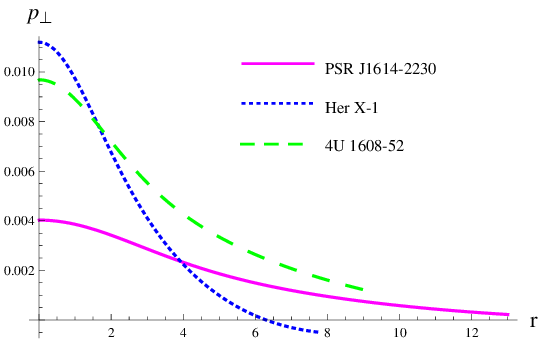,width=0.4\linewidth}\epsfig{file=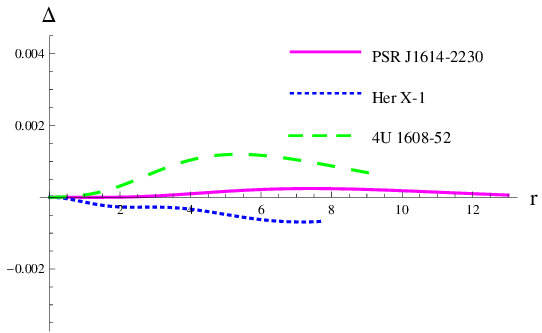,width=0.4\linewidth}
    \caption{Behavior of $\rho,~p_r,~p_\perp$ (in $km^{-2}$) and
        $\Delta$ of anisotropic solution II for $\sigma=0.2$.}
\end{figure}
\begin{figure}\center
    \epsfig{file=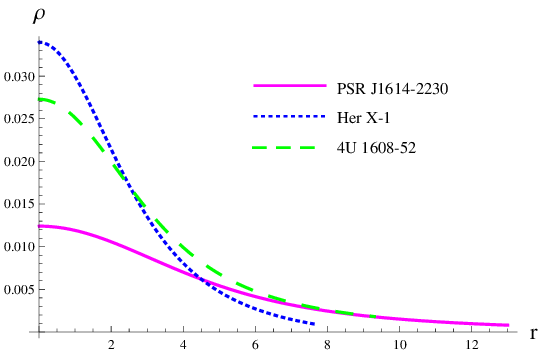,width=0.4\linewidth}\epsfig{file=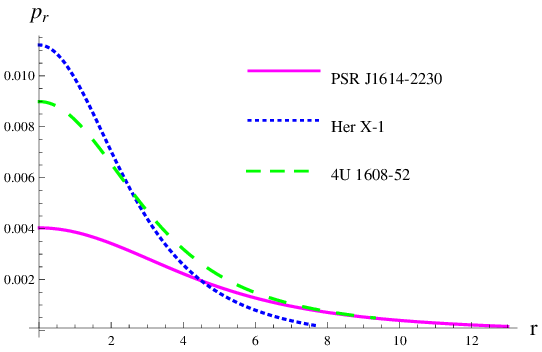,width=0.4\linewidth}
    \epsfig{file=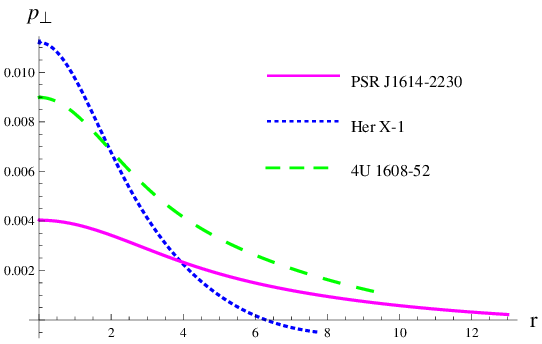,width=0.4\linewidth}\epsfig{file=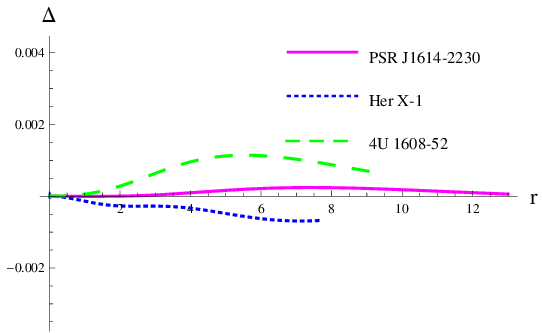,width=0.4\linewidth}
    \caption{Behavior of $\rho,~p_r,~p_\perp$ (in $km^{-2}$) and
        $\Delta$ of anisotropic solution II for $\sigma=0.9$.}
\end{figure}
The plots of radial and temporal metric functions in Figures
\textbf{12} and \textbf{13} are free from singularities with
increasing trend away from the center. Thus, the anisotropic strange
star is well-behaved under the effect of self-interacting scalar
field. Figures \textbf{14} and \textbf{15} indicate that energy
density as well as pressure components are finite and maximum at
$r=0$ with a decreasing trend towards the surface. Further,
tangential pressure corresponding to Her X-I attains negative values
near the boundary. The anisotropy is zero at center and positive throughout
the interior region for PSR J1614-2230 and 4U 1608-52. However, the
model related to Her X-I has negative anisotropy for
$\sigma=0.2,~0.9$. The anisotropic extension is viable as the
parameters governing the state of the distribution satisfy the
energy conditions for lower as well as higher values of the
decoupling parameter (Figures \textbf{16} and \textbf{17}).

\begin{figure}\center
\epsfig{file=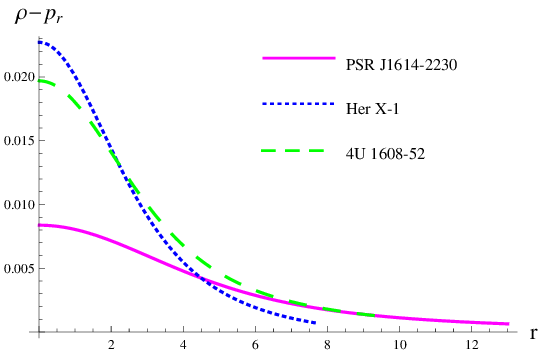,width=0.4\linewidth}\epsfig{file=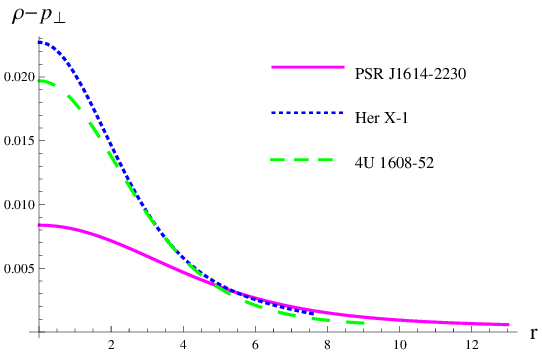,width=0.4\linewidth}
\caption{Graphical representation of DEC (in $km^{-2}$) for extended solution II
for $\sigma=0.2$.}
\end{figure}
\begin{figure}\center
\epsfig{file=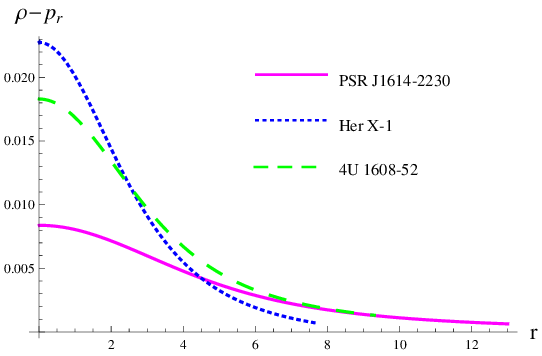,width=0.4\linewidth}\epsfig{file=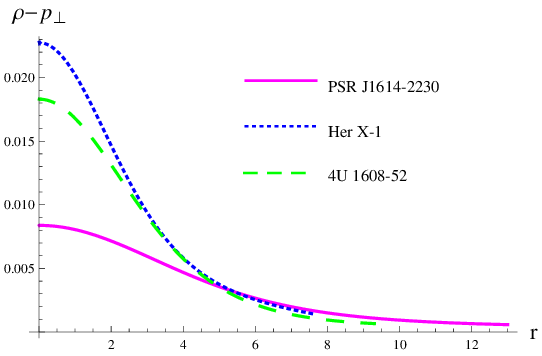,width=0.4\linewidth}
\caption{Graphical representation of DEC (in $km^{-2}$) for extended solution II
for $\sigma=0.9$.}
\end{figure}
\begin{figure}\center
\epsfig{file=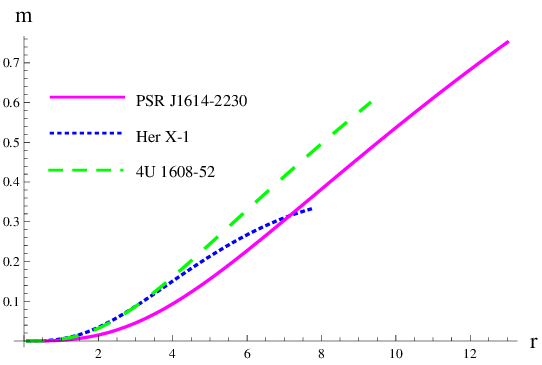,width=0.35\linewidth}\epsfig{file=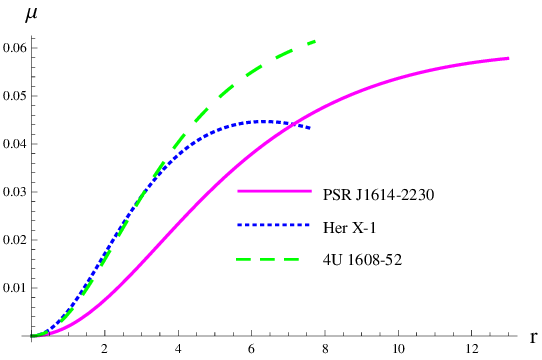,width=0.35\linewidth}
\epsfig{file=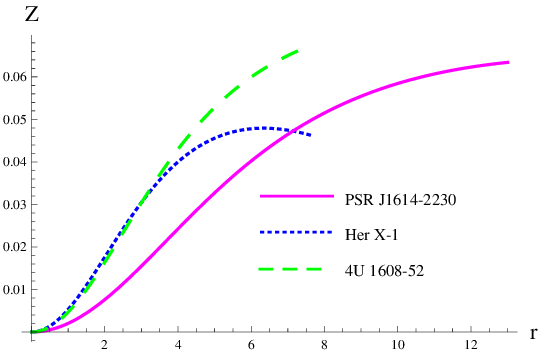,width=0.35\linewidth} \caption{Graphical
representation of $m$ $(M_{\bigodot}$), $\mu$ and $Z_s$ corresponding to anisotropic
solution II for $\sigma=0.2$.}
\end{figure}
\begin{figure}\center
\epsfig{file=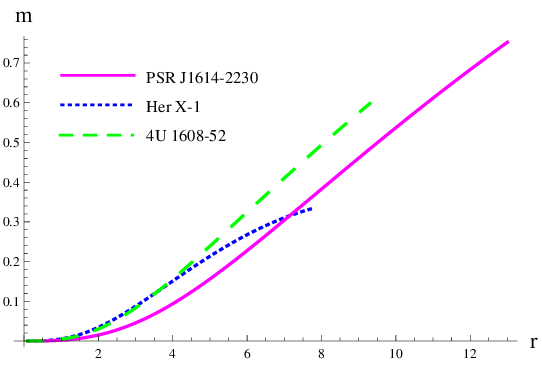,width=0.35\linewidth}\epsfig{file=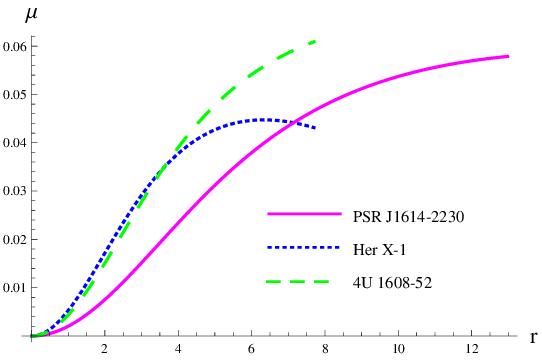,width=0.35\linewidth}
\epsfig{file=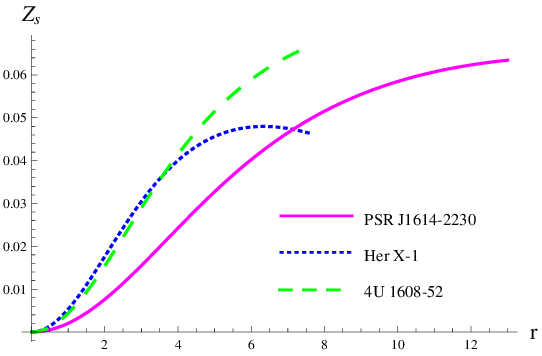,width=0.35\linewidth} \caption{Graphical
representation of  $m$ $(M_{\bigodot}$), $\mu$ and $Z_s$ corresponding to anisotropic
solution II for $\sigma=0.9$.}
\end{figure}
\begin{figure}\center
\epsfig{file=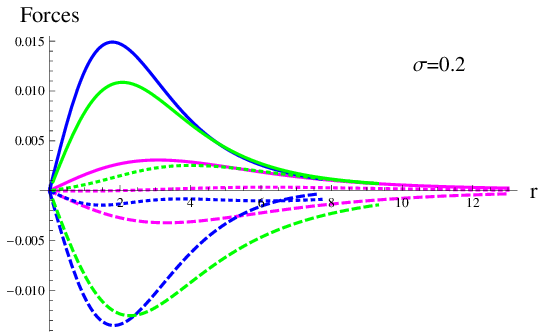,width=0.4\linewidth}\epsfig{file=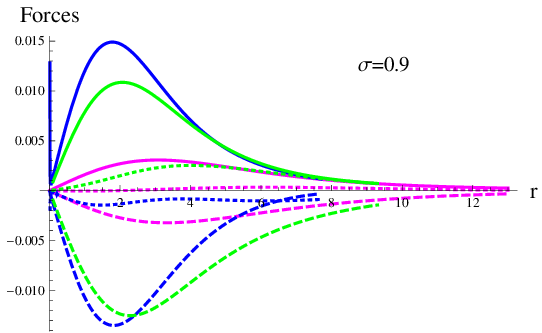,width=0.4\linewidth}
\caption{Graphical representation of $f_h$ (solid), $f_g$ (dashed) and $f_a$
(dotted) for PSR J1614-2230 (pink), Her X-1 (blue) and 4U 1608-52 (green) corresponding to solution II with $\sigma=0.2$ and $0.9$.}
\end{figure}
\begin{figure}\center
\epsfig{file=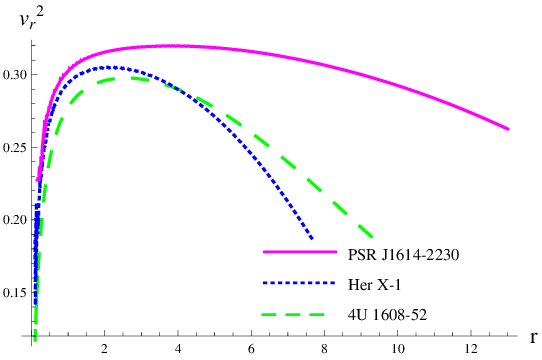,width=0.4\linewidth}\epsfig{file=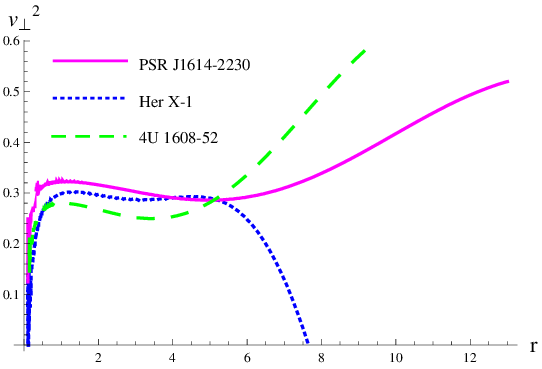,width=0.4\linewidth}
\epsfig{file=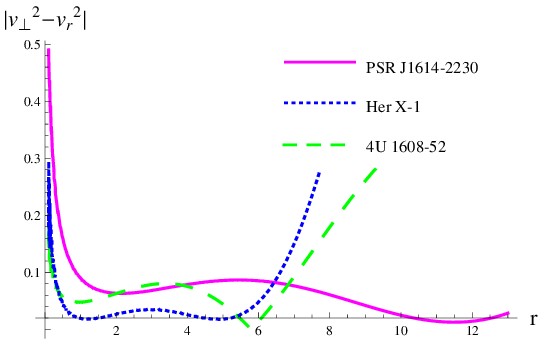,width=0.4\linewidth}\epsfig{file=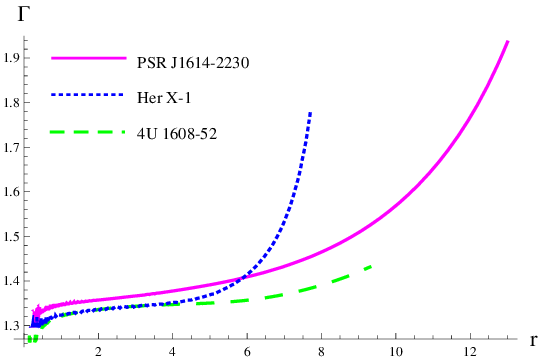,width=0.4\linewidth}
\caption{Graphical representation of radial/tangential velocities,
$|v_\perp^2-v_r^2|$ and adiabatic index of extended solution II for
$\sigma=0.2$.}
\end{figure}
\begin{figure}\center
\epsfig{file=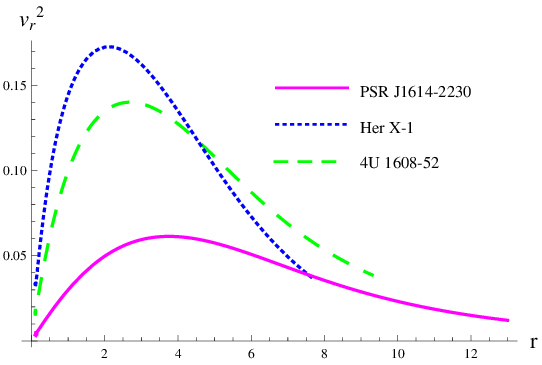,width=0.4\linewidth}\epsfig{file=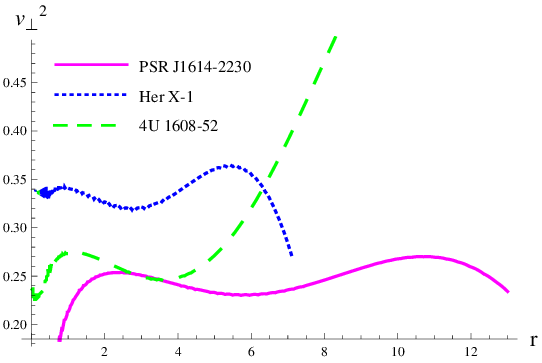,width=0.4\linewidth}
\epsfig{file=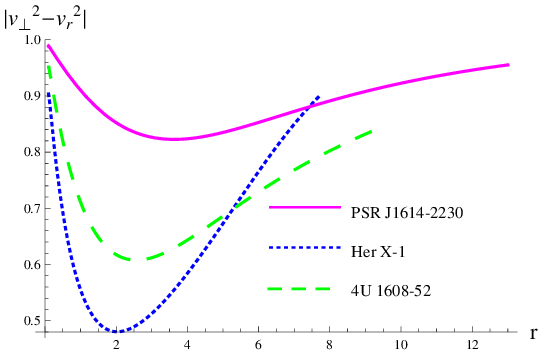,width=0.4\linewidth}\epsfig{file=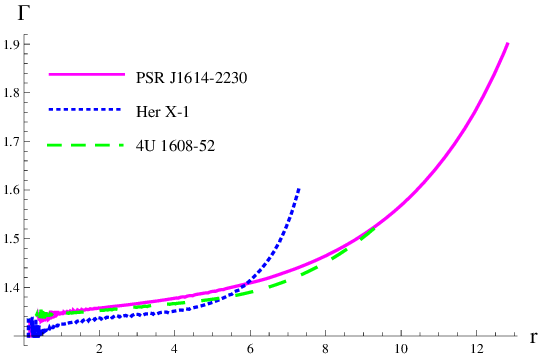,width=0.4\linewidth}
\caption{Graphical representation of radial/tangential velocities,
$|v_\perp^2-v_r^2|$ and adiabatic index of extended solution II for
$\sigma=0.9$.}
\end{figure}

The mass of the anisotropic distribution is obtained by inserting
Eq.(\ref{37}) in (\ref{31}). The compactness factor and redshift
function are shown in Figures \textbf{18} and \textbf{19} for
$\sigma=0.2$ and 0.9, respectively. These functions are consistent
with their respective upper bounds. The hydrostatic, gravitational
and anisotropic forces related to the second setup are plotted in
Figure \textbf{20} for $\sigma=0.2$ and $0.9$. It is noted that
the anisotropic model is in equilibrium as the overall effect of
the forces vanishes. We inspect the anisotropic
stellar model for stability via causality and cracking criteria.
Figures \textbf{21} and \textbf{22} reveal that the components of
sound velocity satisfy the inequalities $0<v_\perp^2<1,~0<v_r^2<1$
and $0<|v_\perp^2-v_r^2|<1$. Thus, the generated anisotropic
extension is stable. Furthermore, the adiabatic index associated
with 4U 1608-52 is below $\frac{4}{3}$ initially but it increases
with an increase in $r$. On the other hand, the adiabatic index
related to the remaining two stars observe the required limit.

\section{Conclusions}

Self-gravitating systems in the universe serve as an effective tool
for comprehending the mechanism of the universe. A strange star is a
compact object that evolves from a neutron star. In this paper, we
have computed anisotropic spherical solutions to analyze salient
properties of quark stars in the context of SBD gravity. For this
purpose, we have extended known isotropic solutions to the
anisotropic domain through the method of gravitational decoupling.
Anisotropy has been produced in the seed matter configuration via an
additional source. The two sources (additional and seed) have been
separated from each other by means of a transformation in the radial
metric component. As a consequence, we have obtained two systems of
differential equations. The first system includes the effects of the
seed distribution only while the influence of the additional source
is restricted to the second set. We have considered Tolman V
spacetime to represent the first array in solution I whereas the
second solution has been generated by imposing the embedding
class-one to generate a line element corresponding to the radial
component $\lambda=\ln\left(\frac{2 c r^2+1+\cosh \left(2 a
r^2+2b\right)}{1+\cosh \left(2 a r^2+2b\right)}\right)$. The system
related to the anisotropic source has been solved by implementing
MIT bag model on matter variables of the anisotropic solution. We
have numerically evaluated the wave equation for
$V(\varsigma)=\frac{1}{2}m_{\varsigma}^2\varsigma^2,~m_{\varsigma}=0.001$
and $\sigma=0.2,0.9$. Finally, the prominent features of the
extended anisotropic models have been inspected for
$\mathcal{B}=60MeV/fm^3$ against different criteria for the stars
Her X-1, PSR J1614-2230 and 4U 1608-52.

The graphical analysis of the metric components has shown that both
stellar models are well-behaved. The state parameters observe a
monotonically decreasing trend towards the surface and attain the
maximum value at $r=0$. Moreover, the anisotropy corresponding to
Her X-1 and PSR J1614-2230 is negative whereas it is positive
throughout the interior region for 4U 1608-52 in solution I. On the
other hand, the anisotropy in the second solution is negative for
Her X-1 only. Thus, radial pressure is less than the transverse
component indicating a presence of repulsive force within the
spherical object corresponding to 4U 1608-52 in solutions I and II.
The extended models are composed of viable normal matter as they are
consistent with the energy conditions. Furthermore, the compactness
and redshift parameters lie below their respective upper bounds. We
have established that the constructed setups are stable as
radial/tangential components of speed agree with causality and
cracking criteria. The adiabatic index of the second anisotropic
model is less than $\frac{4}{3}$ for the star 4U 1608-52 near the
center. However, in all other scenarios, the adiabatic parameter is
greater than $\frac{4}{3}$ which indicates that the static models
hold themselves against inward gravitational pull for the considered
values of the parameters. Thus, the strange star structures
formulated via decoupling are physically realistic and stable in the
presence of a massive scalar field. It is interesting to mention
here that all the obtained results reduce to GR for
$\varsigma=\text{constant}$ and $\omega_{BD}\rightarrow \infty$.

\end{document}